\documentclass[a4paper,11pt]{article}
\usepackage{graphicx}
\usepackage[space]{grffile}
\usepackage{latexsym}
\usepackage{textcomp}
\usepackage{longtable}
\usepackage{tabulary}
\usepackage{booktabs,array,multirow}
\usepackage{amsfonts,amsmath,amssymb}
\usepackage{natbib}
\usepackage{float}
\usepackage{url}
\usepackage{hyperref}
\hypersetup{colorlinks=false,pdfborder={0 0 0}}
\usepackage{etoolbox}
\makeatletter
\usepackage{authblk}
\usepackage[utf8]{inputenc}
\usepackage[english]{babel}
\usepackage[top=1in, bottom=1in, left=1in, right=1in]{geometry}
\title{A Global Tropical Survey of Mid-Tropospheric Cyclones}
\date{}
\author[1,2]{Pradeep Kushwaha \thanks{Corresponding author: Pradeep Kushwaha, pkushwaha@gmail.com}}
\author[1,2]{Jai Sukhatme}
\author[1,2,3]{Ravi S Nanjundiah}
\affil[1]{Centre for Atmospheric and Oceanic Sciences, Indian Institute of Science, Bangalore, 560012, India}
\affil[2]{Divecha Centre for Climate Change, Indian Institute of Science, Bangalore, 560012, India}
\affil[3]{Indian Institute of Tropical Meteorology, Pune, 411008, India}
\begin{document}
%\linenumbers
\maketitle
%\linenumbers
%\clearpage

\begin{abstract}
Mid-Tropospheric Cyclones (MTCs) are moist synoptic systems with distinct mid tropospheric vorticity maxima and weak signatures in the lower troposphere.
Composites and statistics of MTCs over the tropics are constructed and compared with monsoon lows and depressions (together, lower troposphere cyclones; LTCs).
We begin with South Asia, where tracking reveals that MTCs change character during their life, i.e., their track is composed of MTC and LTC phases. The highest MTC-phase density and least motion is over the Arabian Sea, followed by the Bay of Bengal and South China Sea.
A MTC-phase composite shows an east-west tilted warm above deep cold-core temperature anomaly with maximum vorticity at 600 hPa. While the LTC-phase shows a shallow cold-core below 800 hPa and a warm upright temperature anomaly with lower tropospheric vorticity maximum.
Apart from South Asia, systems with similar morphology are observed over the west and central Africa, east and west Pacific in boreal summer. In boreal winter, regions that support MTCs include northern Australia, the southern Indian Ocean, and South Africa.
Relatively, the MTC fraction is higher equatorward, where there is a cross-equatorial low-level jet that advects oppositely signed vorticity. Whereas the LTCs are more prevalent further poleward.
Finally, a histogram of differential vorticity (difference between middle and lower levels) versus the height of peak vorticity for cyclonic centers is bimodal. One peak, around 600 hPa, corresponds to MTCs while the second, around 900 hPa, comes from LTCs. Thus, moist cyclonic systems in the tropics have a natural tendency to reside in either the MTC or LTC category preferentially
\end{abstract}

\section{Introduction}
The Indian summer monsoon rainfall (June-September) exhibits large variations in space and time \citep{LauWaliser}. This space-time variability is a result of various 
interacting phenomena that include the northward propagation of Inter-Tropical Convergence Zone 
\citep[ITCZ,][]{sikka1980maximum}, i.e., the boreal summer intraseasonal oscillation \citep{nanjundiah1992intraseasonal, wang}, influence of the Madden-Julian Oscillation \citep{annamalai2001active}, 
the quasi-biweekly mode
\citep{kb-1976} and monsoon lows \citep[called monsoon depressions, or MDs, if more intense,][]{Mooley}. In particular, monsoon lows and depressions contribute significantly to Central Indian summer monsoon rainfall \citep{HurleyBoos,Hunt,Adames,Vishnu}. These systems and their stronger versions (known as tropical cyclones; TCs) occur throughout the tropics and have an upright vorticity structure with wind and convergence maxima in the lower troposphere. 
Though in comparison to TCs, which have a warm anomaly through the depth of the troposphere \citep{wang201913}, lows and depressions usually show a shallow near surface cold-core and a bimodal potential vorticity structure with one peak near 500 hPa and another near the boundary layer \citep{Boos2015,murthy2019understanding}. 

\noindent In contrast, Western India receives a fair amount of rainfall from moist synoptic disturbances called Middle Troposphere Cyclones \citep[MTCs,][]{carr1977mid,choudhury2018phenomenological}. The first MTC discovered in this region was during the International Indian ocean experiment from 2-10 July 1968 \citep{miller1968iioe}. This particular MTC remained over the northeast Arabian Sea and produced heavy rain that was recorded in many stations along the west coast of India \citep{miller1968iioe}. MTCs are characterized by vorticity and convergence maxima in the middle troposphere and a weak signature in the lower troposphere 
Moreover, they usually also have a pronounced zonal and meridional tilt of vorticity with height \citep{miller1968iioe,carr1977mid}. This tilt is accompanied by a baroclinic warm above cold temperature anomaly. %The MTCs potential vorticity structure mainly unknown as of date.}
 As far as motion characteristics are concerned, MTCs occasionally remain quasi-stationary for several days over the North East Arabian Sea \citep{ramage1966summer} and have proven to be responsible for floods over Western India, especially in the early phase of monsoon \citep{choudhury2018phenomenological}. 
Following \cite{ramage1966summer} who put forth that vorticity export from the Arabian heat low could act as a trigger, several attempts have been made to understand the formation and maintenance of these systems \citep{carr1977mid, mak1983moist, goswami1980role}. 
One of the vexing issues in identifying instabilities associated with the genesis of these systems in this region is, as demonstrated later in the manuscript, that many of them have a nonlocal origin, i.e., they begin as lows or depressions in distant geographical regions and take on features of MTCs later in their life cycle. 

\noindent In addition to the Arabian Sea, there is evidence of MTC occurrence over Southern Indochina and the Bay of Bengal \citep{krishnamurti1970mid}. In the tropics, another example of systems with prominent middle tropospheric maxima and a deep cold-core in the lower troposphere are African Easterly Waves \citep[AEWs;][]{burpee,reed1977structure,jenkins1995cold,russell2020potential}. These waves contribute to the intraseasonal variability associated with the West African monsoon \citep{nicholson2003} and sometimes develop into hurricanes over the Atlantic Ocean \citep{hopsch2010analysis}. 
While barotropic and baroclinic processes are essential in the genesis of these disturbances \citep{thorn-hoskins}, their mid-tropospheric nature could be related to the location of maxima of meridional potential vorticity and heating gradients \citep{hsieh2008instability, thorncroft2008three}. Recent work concerning MTCs over the Arabian Sea and AEWs suggests that stratiform convection plays a role in enhancing their mid-level circulation and vorticity \citep{choudhury2018phenomenological,russell2020african}. Further, case studies of easterly waves have reported prominent mid-tropospheric signatures, though not necessarily a closed cyclonic circulation, near Central America \citep{simpson1967study,ReedR}, the Western Atlantic \citep{Shapiro}, and the Eastern Pacific  \citep{Raymond_ep} regions. 
\noindent Given this background, in this work, we construct a climatology of MTCs --- systems with sustained vorticity maximum in the middle troposphere --- over the global tropics. In fact, as with the recent study of monsoon lows by \cite{HurleyBoos}, we expect that a better and more general understanding of these systems can be attempted with this global tropical (30$^\circ$S to 30$^\circ$N) perspective.
%Hence, we then explore the global tropics ) for such moist mid-tropospheric systems. 
In particular, analysis is carried out in the boreal summer (June to September) and winter (December to March) for 20 years (2000-2019). Along with MTCs, we also keep track of tropical systems with lower troposphere vorticity maxima, which includes monsoon lows, MDs and TCs (together referred to as lower tropospheric cyclones; LTCs), and the relative fraction of these two types of systems (i.e., MTCs and LTCs) is also analyzed.
In essence, here, we address the following questions:
\begin{enumerate}
    \item What are the frequency and statistical characteristics (genesis, lysis, cyclone motion vectors, and track density) of MTCs over South Asia? 
    
    \item What are the differences in the temperature and vorticity profiles of MTCs and LTCs as deduced from a composite that samples systems from all of South Asia? 
    
    \item Are MTCs fundamentally independent of tropical lows (or LTCs)? Or do MTCs and LTCs appear in different stages of the life cycle of tropical cyclonic systems?

    \item What is the global tropical distribution of MTC activity? Are there any common features among the geographical regions that support MTCs? Further, are composites of MTCs from around the tropics similar to those from South Asia?

\end{enumerate}

Details of the data used are presented in the next section. The methodology of detection and tracking of MTCs is presented in the third section. In particular, we provide details on the criteria used to categorize cyclonic centers as MTCs and LTCs and justify the various thresholds used in this process. 
We also introduce metrics such as cyclone center density, the cyclone motion vector, and track density that are used to characterize cyclonic activity in various regions across the globe. The results are presented in section four. Specifically, we first focus on South Asia by tracking systems for sixteen years. Next, we take a global view and present results from the northern and southern hemispheres in the boreal summer and winter, respectively. 
The robustness of results, especially in the context of reanalysis data, is then discussed in the fifth section. Finally, the sixth section summarizes the results and points out questions that this analysis brings to the fore.

%================================================================
                    % DATA SECTION BEGIN
%================================================================
\section{Data}
In this study, we use once daily (12:00 UTC) Modern-Era Retrospective analysis for Research and Applications Version-2 (MERRA-2) reanalysis spanning years 2000-2019 {\citep{gelaro2017modern}}.
The data is available at a regular longitude-by-latitude grid with resolution of $0.625^{\circ} \!\times\! 0.5^{\circ}$ and on 72 hybrid-eta levels from the surface to 0.01 hPa. The MERRA-2 system retains many of the basic features of MERRA \citep{rienecker2008geos}. However, it has several essential updates: aerosol data assimilation, changes to the forecast model, and bias correction using aircraft observations \citep{gelaro2017modern}. 
An overview and assessment of these changes are documented in a special collection of articles devoted to MERRA-2 (\url{https://journals.ametsoc.org/collection/100/Modern-Era-Retrospective-analysis-for-Research-and}). In particular, 
%A recent study by \cite{hodges2017well} 
significant improvements have been noted in MERRA-2, especially year post 2000, in terms of representation of tropical cyclone wind and mean sea-level pressure (MSLP) intensities \citep{hodges2017well}. In fact, MERRA-2 was found to be comparable in many cyclone metrics with NCEP-CFSR and JRA-55 reanalyses. 
Moreover, most modern reanalysis data, including MERRA-2, are also able to capture the outer size and wind structure of tropical cyclones \citep{schenkel2017evaluating}. Keeping this in mind, and to utilize the accuracy of assimilation of the growing observational network \citep{hodges2017well, gelaro2017modern}, we have selected the duration 2000-2019 for our study. 

\noindent Data on 25 vertical levels between 1000 hPa and 100 hPa is used. As the systems, we are studying are greater than 150 km in scale, the data sets are interpolated to a 1.5 degree horizontal rectangular latitude-longitude grid from native resolution prior to application. Such reduced resolution usually helps to filter-out small-scale vorticity centers, whose inclusion introduces errors in tracking features of interest \citep{bengtsson2007tropical}.
Further, for tracking synoptic-scale systems, the use of daily data, rather than finer temporal resolutions, does not affect the nature of results (see Supplementary Material; Figure S1). For a given computational capacity, this also allows for the use of high vertical resolution global data. 
Though finer temporal resolution data is available, recent work by \cite{Vishnu} used 12-hourly MERRA-2 and other reanalyses data for identifying and tracking low-pressure systems during the Indian monsoon season.  In all, \cite{praveen2015relationship} found that the use of daily reanalysis data to track monsoon lows yielded results that are largely similar to those which use finer temporal resolution data \citep{HurleyBoos,Hunt}. 

\noindent Over the past decade, the emergence of high-resolution reanalysis data sets that assimilate vast amounts of satellite and ground-based observations have shown promising results in explorations of the climatology and structure of tropical lows and depressions \citep{HurleyBoos,praveen2015relationship}. While there are issues with under resolution of the lower tropospheric circulation \citep{manning2007evolution,hodges2017well}, these products have been successfully used to understand rainfall distribution, dynamic \&  thermodynamic details, and the propagation of tropical cyclonic systems \citep{Hunt,Boos2015,sorland2015dynamic,hunt2016spatiotemporal,murthy2019understanding}. 
Here, our objective is to understand the structure of tropical systems with vorticity maxima in the middle or lower troposphere. We believe this is feasible as the studies mentioned above have shown that horizontal wind and vorticity fields (on synoptic scales) are faithfully captured by most of the latest generation reanalysis data.
In fact, it has been possible to differentiate between systems with different thermal structures (cold versus warm core) using reanalysis data based on vorticity or geopotential height fields \citep{hart2003cyclone, bengtsson2007tropical,sorland2015dynamic}.

%================================================================
%                           METHOD BEGINS
%================================================================
\section{Parameter selection, identification and tracking of systems}

The identification of systems follows conventional cyclone detection methods \citep{sinclair1994objective, lambert1996intense}. The distinction between MTCs and LTCs requires an additional crucial step of differentiating between systems based on the vertical position (lower or middle troposphere) of their relative vorticity maximum. The procedure starts with checking each grid point of the 600 hPa geopotential height field to locate local minima. We choose the 600 hPa surface as MTCs \citep{carr1977mid,choudhury2018phenomenological}, monsoon lows \citep{HurleyBoos,Hunt} and tropical cyclones \citep{wang201913} have a signature at this level.
This detection of minima is fully automated. Once cyclonic centers are identified, they are stored with properties such as their latitude, longitude, level of vorticity maximum ($P_{\xi}$), mean middle (650-500 hPa) and lower layer (1000-850 hPa) vorticity ($\xi_{m}$ and $\xi_{l}$) and mean middle layer-specific humidity ($Q_{m}$). These properties are averaged over a $4^{\circ}\times 4^{\circ}$ latitude-longitude box in the storm. We also compute $\delta \xi_p = \xi_{m}-\xi_{l}$, i.e., the difference between middle and lower layer vorticity. The area-average is used to ensure that the sign of $\delta \xi_p$ is not local, moreover, this is the typical size of the core of MTCs at 600 hPa within relative vorticity contour of $\xi>2$ $\times 10^{-5}s^{-1}$ \citep{carr1977mid, miller1968iioe}.
Then, the classification of systems either as middle troposphere cyclonic center or lower troposphere cyclonic centers is based on the following constraints:

\begin{enumerate}           
\item For MTCs, relative vorticity maximum must be in the middle troposphere (i.e., $650\geq P_{\xi}\geq 500 $ hPa); the mean middle and lower layer vorticity must be positive (i.e., $ \xi_{m}>0$, $\xi_{l}>0$) and mean middle layer vorticity must exceed that of the lower layer --- $\delta \xi_{p} > 1.5$ $\times 10^{-5} s^{-1}$. 
  
\item For LTCs, relative vorticity maximum must be in the lower troposphere (i.e., $1000\geq P_{\xi}  \geq700$ hPa); the mean lower and middle level vorticity must be positive (i.e., $ \xi_{m}>0$, $\xi_{l}>0$) and mean lower layer vorticity must either exceed that of the middle layer, or be comparable to it, hence --- $\delta \xi_{p} < 0.5$ $\times 10^{-5} s^{-1}$.
 
\item Further, only moist and intense systems are considered by imposing the cutoffs $Q_{m}>2$ g/kg and $\xi_{m}  > 1.5\times 10^{-5} s^{-1}$, respectively. 
\end{enumerate}

\noindent The choice of these thresholds for layers, intensity and moisture content are discussed next.

\subsection{Selection of parameters}

We begin by presenting the typical structure of MTCs in the Indian region (where they have been previously identified) and contrast it with other tropical systems such as monsoon depressions or lows. These features will be used to frame constraints so as filter out particular types of systems.
Apart from a few case studies \citep{miller1968iioe, krishnamurti1970mid}, to our knowledge, the only MTC events cataloged are by the IMD during the years 1998-2008 over the Northeast Arabian Sea \citep{choudhury2018phenomenological}. Here, we take advantage of these 35 middle troposphere circulation events to estimate the structure of MTCs. Noting that Arabian Sea MTCs usually co-exist with MDs over the Bay of Bengal \citep{carr1977mid,choudhury2018phenomenological}, we extend the region of inspection to cover both these basins (i.e., $5^{\circ}$N to $25^{\circ}$N and $50^{\circ}$N to $95^{\circ}$E) to sample a spectrum of monsoon systems. Minima of the 600 hPa geopotential in this domain throughout these 35 events (i.e., 294 days) are identified and results in 725 %\textcolor{red}{(Some logical error fixed and number is modified)} 
non-topographic\footnote{We reject any system as topographic if it has missing  vorticity values at the center and/or if any velocity component has missing values within four degrees east, west, north and south of the center at 600 hPa.} strong ($\xi_{m}  > 1.5\times 10^{-5} s^{-1}$) and moist ($Q_{m}>2$ g/kg) cyclonic centers. The threshold of $\xi_{m}  > 1.5\times 10^{-5} s^{-1}$ is used to filter out weaker vorticity centers \citep{Hunt,Boos2015}.

\begin{figure*}    
\centering
\includegraphics[trim=0 0 0 0, clip,height = 1\textwidth,width = 0.6\textwidth, angle =90, clip]{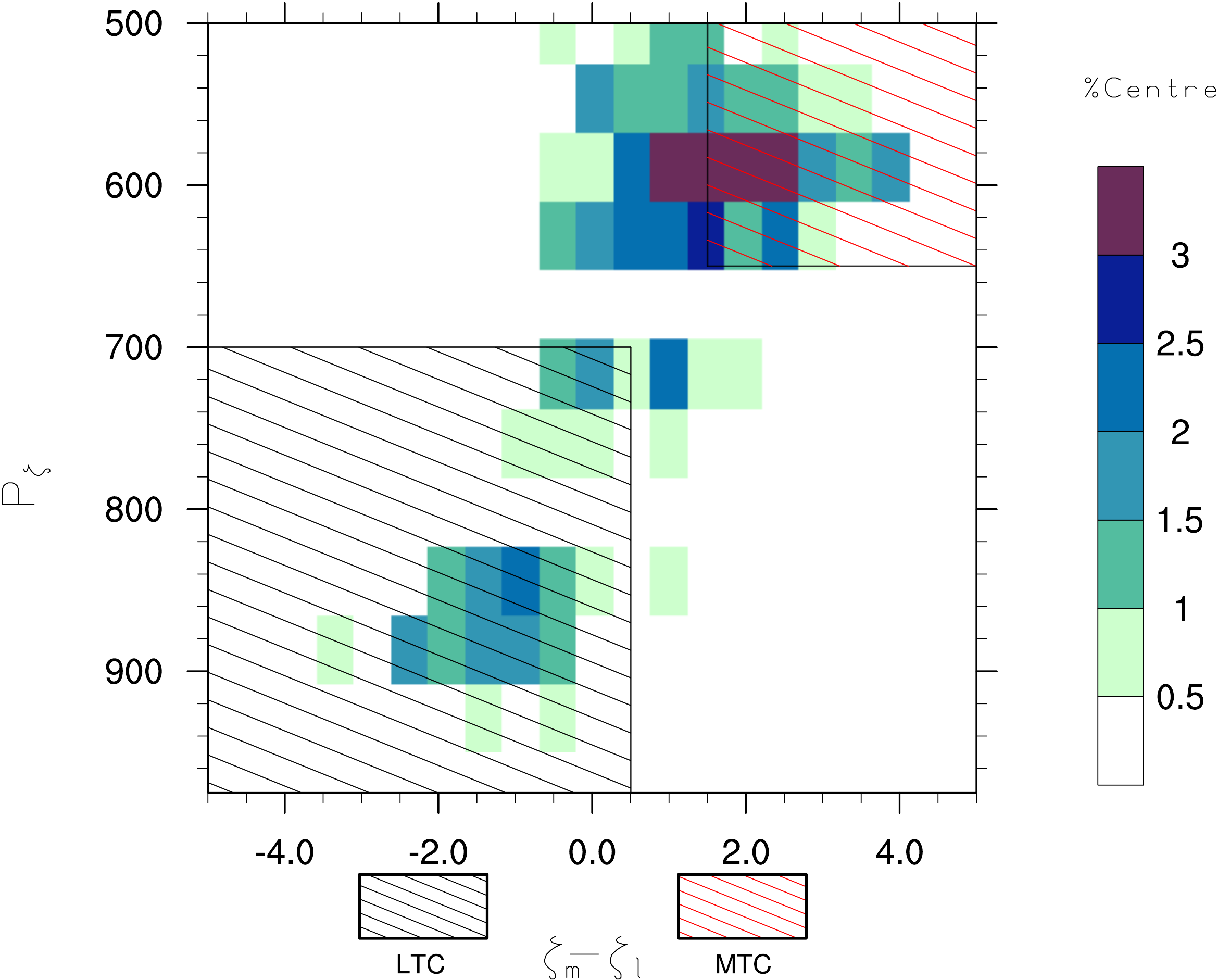}
\caption{Joint probability distribution of $\delta \xi_p = \xi_m - \xi_l$ ($\times 10^{-5}$ s$^{-1}$) and the level of maximum relative vorticity ($P_{\xi}$, in hPa) for $725$ strong \& moist cyclonic centers detected during the dates of 35 IMD MTC events over the region $5^{\circ}$N-$25^{\circ}$N, $50^{\circ}$E-$95^{\circ}$E.
 Red (black) hatching indicates bounds of MTCs (LTCs).}
\label{fig:FIG1}
\end{figure*}

\noindent The joint probability distribution of $\delta\xi_{p}$ and level of maximum relative vorticity ($P_{\xi}$) of these 725 systems is shown in Figure~\ref{fig:FIG1}. The correlation between differential vorticity and the level of maximum vorticity is $-0.7$. 
This implies that the more substantial the positive/negative difference between middle and lower layer vorticity (i.e., $\delta\xi_{p}$), larger is the possibility that the level of vorticity maximum situated in higher/lower levels and vice versa. %(as discussed for ideal curves in Figure~\ref{fig:MD_MT_TC_PROFILE-crop}e, f). 
It is clear from Figure~\ref{fig:FIG1} that there are two peaks in the locations of the maximum vorticity, namely, between 500-700 hPa and 700-1000 hPa. Further, there is a fair amount of spread in $\delta\xi_{P}$ within both the middle (500-700 hPa) and lower level (700-1000 hPa) peaks. The natural emergence of these two peaks lends credence to studying mid and lower tropospheric systems as two entities. Of course, during its evolution, a system can shift from one category to the other.

\begin{figure*}
\centering
\includegraphics[trim=0 0 0 0, clip,height = 0.6\textwidth,width = 0.7\textwidth, angle =0, clip]{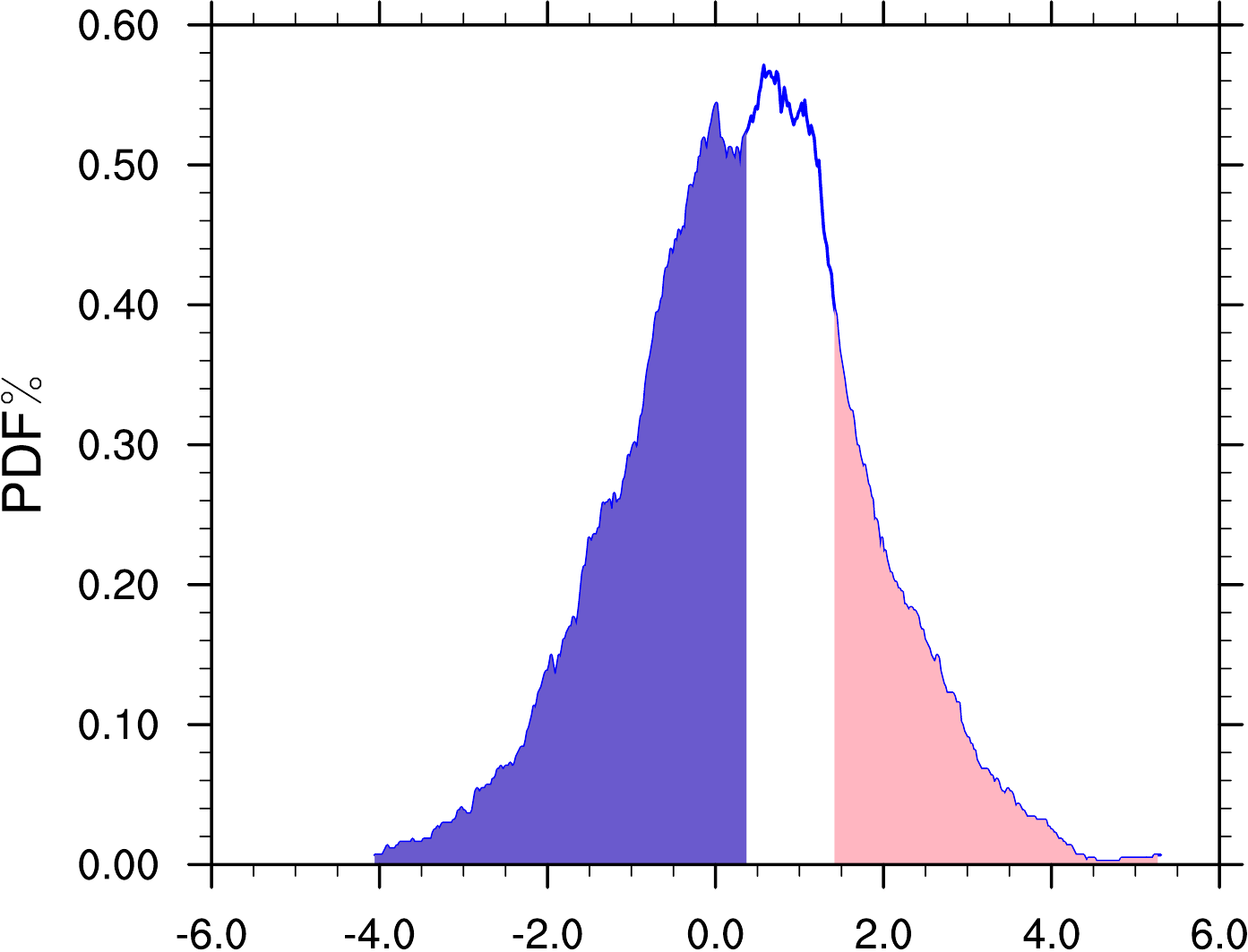}
\caption{Probability distribution of differential vorticity ($\delta \xi_p$, $\times 10^{-5}$ s$^{-1}$) for the same cyclonic centers as in Figure \ref{fig:FIG1}. Light blue shaded region denotes lower 50th percentile (i.e., below $0.5 \times 10^{-5}$ s$^{-1}$), light pink region denotes upper 20th percentile which (i.e., above $1.5 \times 10^{-5}$ s$^{-1}$).} 
\label{fig:FIG2}
\end{figure*}

\noindent Taking these features into account, as shown in Figure \ref{fig:FIG2}, we have selected upper $20^{\textrm{th}}$ percentile of  $\delta\xi_{p}$, i.e., $1.5 \times 10^{-5} s^{-1}$  as a threshold for MTCs and the lower $50th$ percentile of $\delta\xi_{p}$, i.e., $0.5 \times 10^{-5} s^{-1}$ as a threshold for non-MTCs. The choice of these particular percentiles is of course debatable, indeed stricter criteria separate the MTC and non-MTC categories more clearly but at the expense of reducing the number of systems sampled. We have settled on these specific numbers as they serve the purpose of differentiating between the two classes of systems.
Further, we have the constraints $500 \leq P_{\xi} \leq 650$ for MTCs and $700 \leq P_{\xi} \leq 1000$ for non-MTCs.
As is seen from Figure~\ref{fig:FIG1}, most of the systems that comprise this non-MTC category have a vorticity maximum in the lower troposphere; thus, we refer to them as lower tropospheric cyclones (LTCs). Note that this excludes heat lows, which are generally dry and confined below 700 hPa \citep{spengler2008dynamics}. In fact, LTCs include monsoon lows, depressions and tropical cyclones. We remind the reader that we are detecting systems with lower tropospheric maxima to account for the fact that cyclones can change the character during their evolution and also to compare the relative fraction of MTCs to LTCs in a given region. 

\noindent Note that, even though MTCs show vorticity maximum between 700-500 hPa (Figure~\ref{fig:FIG1}), we use a slightly stricter criterion of 650 hPa as a lower bound for $P_\xi$ to avoid the cyclonic shear corresponding to the African Easterly Jet which has maximum intensity around 700 hPa \citep{burpee,reed1977structure}. The sensitivity of $\delta \xi_{p}$ to different choices of layers (Supplementary Material; Figure S2) shows that 
this does not affect the form of the PDF of $\delta P_{\xi}$, though, as expected, this influences the number of MTCs and LTCs.
In summary, the regions defined by the thresholds corresponding to MTCs and LTCs are shown by red and black shading in Figure~\ref{fig:FIG1}, respectively. 

\noindent As we are interested in moist systems, in addition to the  constraint on level of vorticity maximum ($P_{\xi}$) and $\delta P_\xi$, we also require that the geopotential minima detected are moist, i.e., we impose $Q_{m}>2$ g/kg \citep[guided by observations of precipitating systems at 500 hPa,][]{holloway2009moisture}. 
Together, these criteria ensure that the systems considered have a cyclonic circulation and are strong and moist. A verification that our parameters can differentiate between MTCs and LTCs is presented in the Supplementary Material (Figure S3). The sensitivity of the number of systems detected to variations in vorticity magnitude and moisture thresholds is also shown in the Supplementary Material (Table S1).
It is important to note that these thresholds should be treated as tuning parameters and might need adjustment for different data sets, especially with different horizontal resolutions. 

\subsection{Tracking of MTCs}
Following \cite{grigoriev2000innovative}, given its accuracy \citep{gulev2000synoptic}, a manual interactive controlled animation method for tracking the MTCs was implemented. 
This approach is somewhat similar to manual tracking of mesoscale convective systems \citep{davis2002detection}. An automated method for tracking is not employed as there is no manual reference for middle tropospheric systems. Indeed, even very recent uses of automated tracking, see for example, \cite{Vishnu}, require a reference of manually identified and tracked systems for training before the algorithms can be applied on reanalysis fields. In fact, our catalog of manually identified tracks of MTCs can serve as a reference to train automated tracking methods.
The tracking method can be summarized as follows:

\begin{enumerate}

\item Daily maps of the 600 hPa geopotential height and streamlines are examined for local minima. The minima are identified automatically and constitute the detected centers of cyclonic systems. 

  \item These detected centers are labeled with properties including their latitude-longitude location, $\xi_{m}$ and $\delta\xi_{p}$. In all, maps are examined for 1952 days (122$ \times$ 16 = 1952; the number of days in a season $\times$ number of years). This allocation of properties is also fully automated.

   \item A controlled animation of the above maps from one day to another is used to assign tracks for each detected center. During the process, the properties of each center along the tracks are collected manually and digitized.
  
   \item The first occurrence of a MTC within the domain is considered as its genesis location provided $\xi_{m}>1.5 \times 10^{-5} s^{-1}$ for at least two consecutive days and $\xi_{m}-\xi_{l}>1.5 \times 10^{-5} s^{-1}$ for a minimum of two days within its life cycle. 
  
    \item Tracks are terminated when the cyclone shows $\xi_{m}<1.5 \times 10^{-5} s^{-1}$ for two successive days. %\sout{or if the system moves out of the domain}. 
    Further, only systems that last for more than 3 days are considered in the study. 
    
\end{enumerate}

\noindent In essence, the first step is to detect geopotential minimum (system center), then the second step is to attach properties such as $\xi_m$ and $\delta \xi_p$ to each minimum. These two steps are fully automatic. Once the detection and attaching of properties are completed, then manual animation follows. During the animation, conditions 4 and 5 (above) are checked for each center.  While these conditions are satisfied, the track is progressively updated and stored.  

\subsection{Cyclone Metrics}
 Two measures of cyclone activity namely \textit{cyclone center density} and \textit{track density}  \citep{paciorek2002multiple} are used in this study. The cyclone center density counts the number of cyclonic centers within a specific area in a given time duration. When a sufficiently large number of cyclones from multiple years are considered this measure is proportional to the probability of cyclone occurrence in an area \citep{sinclair1994objective}. For computation, we use $8^{\circ}$ square patches 
 surrounding each grid point which leads to a smoothing of patterns and ensures continuity of variables in the resulting maps. 
  
\noindent The second measure, track density, counts the number of cyclones passing within an area centered over each grid point. This measure needs the formation of tracks from detected centers. If derived from sufficiently long data sets, track density is proportional to the probability of a cyclone passing in the vicinity of a grid point \citep{sinclair1994objective}.
Further, each track is interpolated so that at least one track point appears in every grid cell at 1.5$^\circ$ resolution. Finally, the formation of cyclone tracks also allows for the computation of velocity components at each track point and provides the {\it mean cyclone motion vector}.

\section{Results}

\subsection{MTCs over South Asia}

As discussed in the methods section, tracking of middle level (600 hPa) minima over South Asia ($0^{\circ}$N-$30^{\circ}$N, $50^{\circ}$E-$140^{\circ}$E) is carried out manually. We begin our analysis with South Asia because MTCs are frequently reported over the Arabian sea \citep{carr1977mid,choudhury2018phenomenological} and, to some extent, over the Bay of Bengal and the South China Sea \citep{krishnamurti1970mid}.
Apart from an understanding of the regions of genesis, lysis, and propagation of systems, tracking information over South Asia is utilized as a guide to search for systems with similar characteristics in other parts of the tropics.
As mentioned, during tracking, we only consider systems that last for at least three days and show $\delta \xi_p>1.5 \times 10^{-5} s^{-1}$for at least two days. These constraints ensure that the tracked systems are long-lived and allow for the possibility that MTCs can change character in their life cycle. Indeed, we find that most systems identified as MTCs show mid-tropospheric vorticity maxima and lower tropospheric vorticity maxima in different stages of their life cycle.
Thus, the notation we have adopted from here on is that, all systems that show  $\delta \xi_p > 1.5 \times 10^{-5} s^{-1}$ at least for two days are MTCs; the specific days when they show $\delta \xi_p > 1.5 \times 10^{-5} s^{-1}$ is referred to as their MTC-phase and days when $\delta \xi_p < 0.5 \times 10^{-5} s^{-1}$ is called their LTC-phase. Due to the shift of vorticity maxima among levels, we refrained from imposing constraints on $P_{\xi}$ during tracking.  
We have verified that this relaxation does not alter results qualitatively (Supplementary Material; Figure S4); however, as expected, there is an effect on the actual number of MTC centers. 

\noindent In all 261 MTCs are tracked, which contain 2414 track points. Out of these, 1109 track points satisfy MTC-phase criteria and 807 track points LTC phase. Note that, here too, the distribution of $P_\xi$ and $\delta \xi_p$ is bimodal. Our parameters focus attention on the two peaks as we aim to differentiate between the properties of middle and lower tropospheric systems. Hence, quite naturally, there are a fair number of ``intermediate" systems (about 20\%) that do not fall in either category.
These numbers indicate that mid-tropospheric vorticity maxima are not uncommon over South Asia. Specifically, we have 261 such systems in 16 years, 2000 to 2015, from June through September, which implies about 3--4 systems per month. Moreover, systems identified as MTCs change the character and exhibit lower tropospheric maxima during a significant portion of their life cycle. In other words, the MTC-phase does not occur independently from the LTC-phase, and there is conversion from one to the other as systems evolve.

\begin{figure*}
\centering
\includegraphics[trim=0 0 0 0, clip,height = 0.8\textwidth,width = 0.7\textwidth, angle =0, clip]{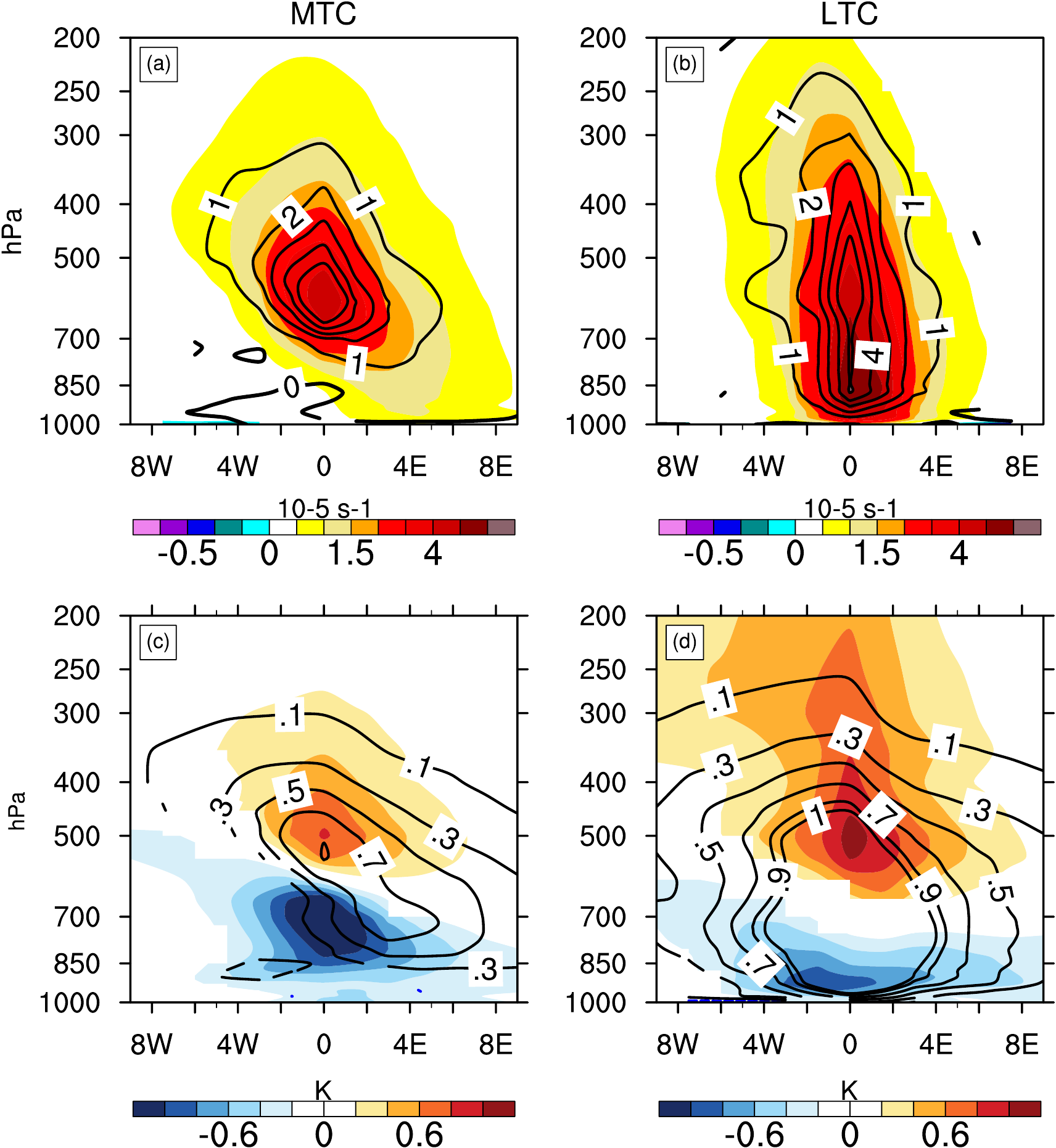}
\caption{South Asian composites from 1107 MTC-phase track points (a) Relative vorticity and PV anomaly in contours (c) Temperature anomaly and specific humidity anomaly contours. In both (a) and (c) positive contours are solid and negative are dashed. Panels (b), (d) are the same quantities but constructed from 807 LTC-phase track points. Unit of relative vorticity is $10^{-5}$s$^{-1}$; Units of PV are $10^{-1}$ PVU; Units of specific humidity are g/kg ; anomalies are from respective daily climatology. Results are only shown if they are significant at 99\% confidence. }
\label{fig:FIG3}
\end{figure*}

\noindent Composites of relative vorticity, potential vorticity (PV), temperature, and specific humidity anomalies of MTC and LTC phases are shown in Figure~\ref{fig:FIG3}. This large sample size provides a robust view of the middle and lower tropospheric phases over South Asia. In particular, and as expected, the MTC (LTC) phase shows a localized vorticity maximum in the middle (lower) troposphere, which is consistent with the systems over the Arabian Sea and the Bay of Bengal (Figure S3). Further, the LTC-phase composite is stronger than MTC-phase in terms of relative vorticity; however, both phases are in gradient wind balance (mean Rossby number $\approx$ 1). The temperature structure depicts warm (cold) anomalies above (below) the level of the MTC center (around 600 hPa). LTCs, on the other hand, shows a warm anomaly from the middle to upper troposphere, and a negative anomaly is squeezed below 800 hPa. The westward tilt of MTCs is apparent in the vorticity and specific humidity anomaly structure, while LTCs do not show such a pronounced tilt with height.
The PV anomaly of MTCs is unimodal with a maximum localized in the middle troposphere, while that of LTCs extends down to the surface and has a bimodal character. Specifically, similar to monsoon lows \citep{Boos2015,Hunt} it shows one maximum in the middle troposphere and another in the lower troposphere. The confinement of MTCs PV in the middle troposphere may be related to the high fraction of stratiform heating \citep{choudhury2018phenomenological,russell2020potential}, while the LTC PV is likely affected by significant deep convective heating \citep{murthy2019understanding}. 

\subsubsection{Track density, genesis density and cyclone motion vector}
\begin{figure*}
\centering
\includegraphics[trim=0 0 0 0, clip,height = 1\textwidth,width = 0.9\textwidth, angle =0, clip]{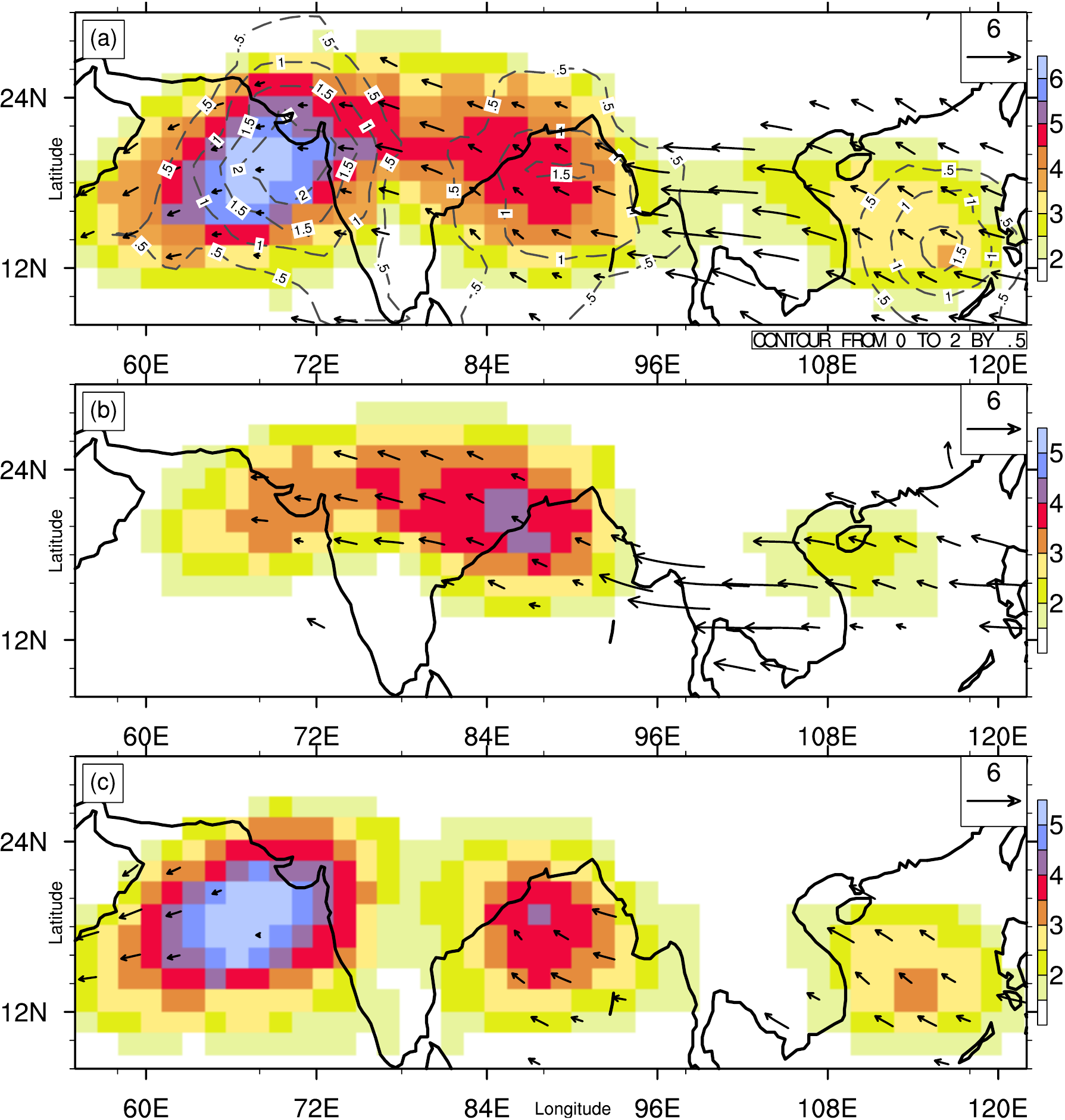}
\caption{(a) Overall track density (color shaded) and genesis density (contours) of 261 MTCs over South Asia. Arrows denote the cyclone motion vectors, arrow length is proportional the speed of propagation in m/s. The vectors are shown only if any of the wind components are significantly different from zero at 99\% confidence under a two-tailed t-test. (b) Same as (a), except calculation is done for each track segment if it is in the LTC-phase (c) is same as (a) except results are for the MTC-phase.}
\label{fig:FIG4}
\end{figure*}

The track density (color shading), genesis density (contour lines), and motion vectors for all systems identified at MTCs and their LTC and MTC-phases are shown in Figure~\ref{fig:FIG4}a, b, c, respectively. The highest genesis density of MTCs (dashed lines), Figure~\ref{fig:FIG4}a, is found over the Arabian Sea ($18^{\circ}$N-$22^{\circ}$N, $68^{\circ}$E-$70^{\circ}$E) followed by Bay of Bengal ($18^{\circ}$N-$22^{\circ}$N, $85^{\circ}$E-$90^{\circ}$E) and South China Sea ($16^{\circ}$N-$22^{\circ}$N, $110^{\circ}$E-$130^{\circ}$E). The genesis maxima over the Arabian Sea is located around $18^{\circ}$N-$22^{\circ}$N, $65^{\circ}$E-$72^{\circ}$E, and is a little southward, $16^{\circ}$N-$20^{\circ}$N, $85^{\circ}$E-$90^{\circ}$E, over the Bay of Bengal.  The genesis maximum over the South China Sea is found near $12^{\circ}$N-$14^{\circ}$N and $110^{\circ}$E-$130^{\circ}$E, which is the most southward and more concentrated in comparison with the other two basins. These locations align well with the reported genesis maximum of low-pressure monsoon systems over the Bay of Bengal and the South China Sea. 
The genesis density over the north Bay of Bengal (1-2 per summer, contour lines) is comparatively less than those reported in previous studies \citep[5-10 per summer,][]{HurleyBoos}. This is primarily because we have considered only systems that show $\delta \xi_p > 1.5 \times 10^{-5} s^{-1}$ at least for two days. On the other hand, the genesis density over the Arabian Sea, despite only considering MTCs, is significantly larger than previous reports. As MTCs show a robust signature in the middle troposphere and many have little or no near-surface signature, this difference is due to our use of the 600 hPa geopotential. In contrast, earlier work used sea level pressure or low-level vorticity as an identifying marker \citep{HurleyBoos}.  

\noindent The track density in Figure~\ref{fig:FIG4} is presented in three parts, for the full track, for the LTC-phase
and for the MTC-phase.
Similar to genesis, three maxima in total, LTC-phase and MTC-phase track density (Figure~\ref{fig:FIG4}a,b,c) are evident in the Northeast Arabian Sea, North Bay of Bengal, and the South China Sea. The Arabian Sea shows the highest total and MTC-phase density while the LTC-phase track density peaks inland and over the head of Bay of Bengal. A comparison of the total, MTC, and LTC-phases suggests that most of the Arabian Sea's total density is contributed from the MTC-phase. While over North Bay of Bengal and inland, the total density is mainly from the LTC-phase. The geographical location of the MTC-phase maximum is qualitatively consistent with the previous literature, which reports MTCs primarily over the North-East Arabian Sea \citep{carr1977mid,choudhury2018phenomenological}. This cyclonic activity matches with enhanced regions of monsoon synoptic activity \citep{HurleyBoos}, indicating that MTCs are found in regions where Indian summer monsoon lows occur. A notable difference is that MTCs show relatively more activity where monsoon depressions and lows show less activity and vice versa. For instance, over the Arabian Sea, MTCs show the highest frequency while monsoon lows have only a modest presence \citep[comparing Figure~\ref{fig:FIG4} with Fig-2 of][]{Boos2015}.
Comparing LTC and MTC phases (Figure~\ref{fig:FIG4}), it appears that the MTC-phase occurs with relatively higher density equatorward compared to the LTC-phase. Together, it is likely that, in their life cycle, middle tropospheric cyclones show LTC and MTC phases; however, there is a preferred range of latitudes for the respective phases.

\noindent From the cyclone motion vectors (only shown if they are significantly different from zero at 99\% confidence with two-tailed t-test) in Figure \ref{fig:FIG4}, cross basin motion from the Bay of Bengal to the Arabian Sea and from the South China Sea to the Bay of Bengal is evident. This is consistent with \cite{carr1977mid}, who in a survey of satellite images found that some monsoon depressions in the Bay of Bengal convert into MTCs as they move westward to the Arabian Sea.
This cross basin propagation, followed by a conversion from LTC-phase to MTC-phase, makes the origin of many of these systems nonlocal. Specifically, an MTC over the Arabian Sea could very well have originated as a depression in the Bay of Bengal.
Further, cyclone motion vectors of MTCs show a northwest orientation over the Bay of Bengal and the South China Sea with fairly large magnitude. However, we see relatively slow westward movement over the western Arabian Sea. In fact, over the northeast Arabian Sea, both LTCs and MTCs do not show a preferred statistically significant direction of motion which is consistent with the observation of their quasi-stationary behavior in this region \citep{miller1968iioe,carr1977mid}.

\subsubsection{Genesis and intensity distributions}
\begin{figure*}
\centering
\includegraphics[trim=0 0 0 0, clip,height = 1.0\textwidth,width = 1\textwidth, angle =0, clip]{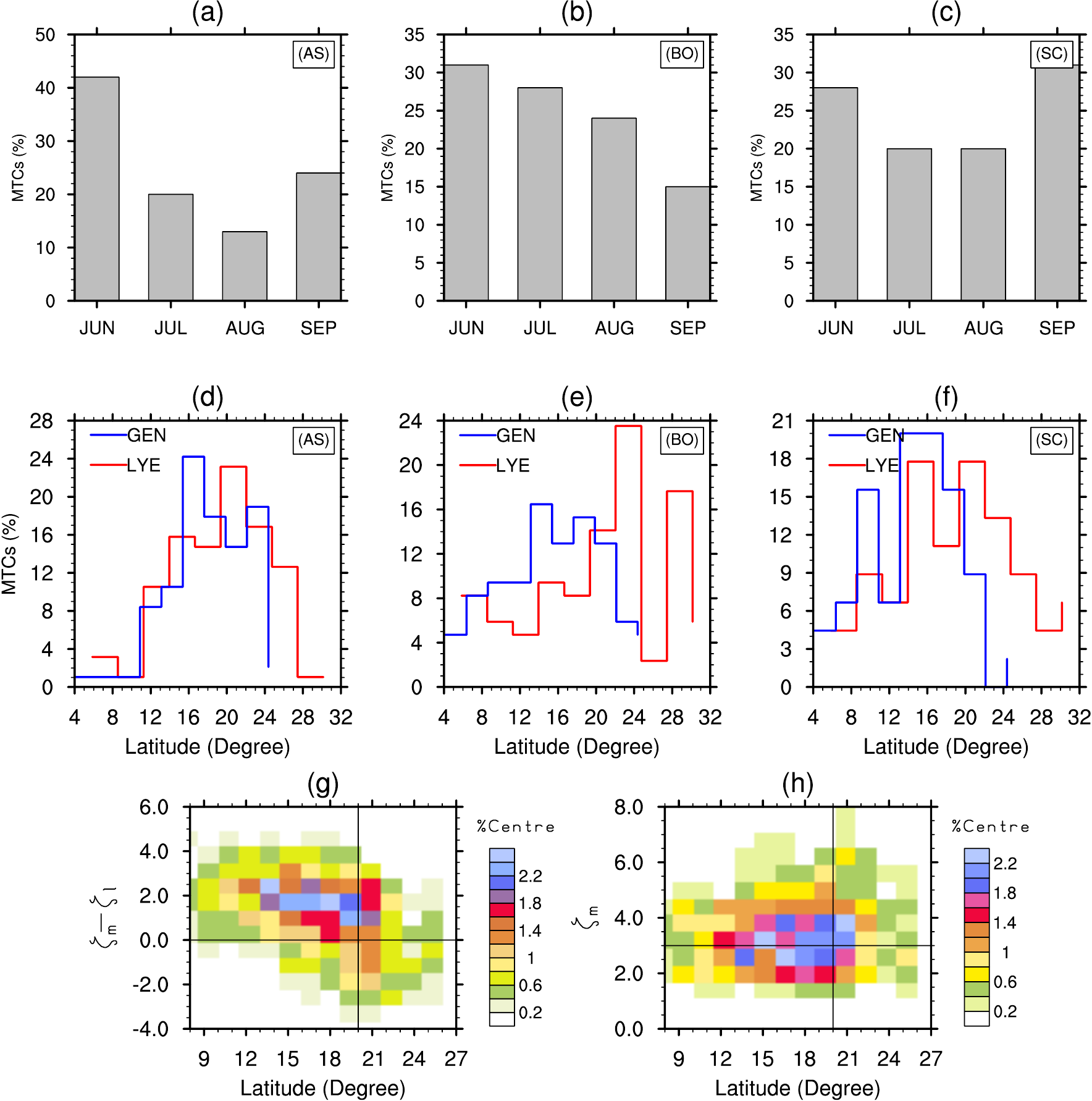}
\caption{Panels (a), (b), (c) show the monthly genesis probability (in percent) of 261 South Asian MTCs over the Arabian Sea (AS~:$50-75^{\circ}$E, 0-30$^{\circ}$N), Bay of Bengal (BO~:$75-100^{\circ}$E, 0-30$^{\circ}$N) and South China Sea (SC~:$100-120^{\circ}$E, 0-30$^{\circ}$N). Panels (d), (e), (f) show the genesis (blue) lysis (red) probability distribution with latitude for the three regions. (g) and (h) show the joint probability distribution of $\delta \xi_{p}$ and $\xi_{m}$ respectively with respect to latitude for all 261 tracks.}
\label{fig:FIG5}
\end{figure*}

Figure~\ref{fig:FIG5}a, b and c shows the monthly probability of occurrence of MTCs over the three identified basins: Arabian Sea ($0^{\circ}$N-$30^{\circ}$N, $50^{\circ}$E-$75^{\circ}$E), Bay of Bengal ($0^{\circ}$N-$30^{\circ}$N, $75^{\circ}$E-$100^{\circ}$E) and South China Sea ($0^{\circ}$N-$30^{\circ}$N, $100^{\circ}$E-$120^{\circ}$E), respectively. Arabian Sea MTC genesis peaks in June (42\%) then reduces during July (20\%) to August (13\%) but increases in the late part of the monsoon during September (23\%).
The high probability of occurrence of MTCs in the early phase of monsoon in this region has been linked with the export of middle-level vorticity from the heat low over Northwest India \citep{ramage1966summer,carr1977mid}, and these systems are likely to be locally born \citep{goswami1980role}.
Over the Bay of Bengal, the highest probability of MTC occurrence is also in the first half of the monsoon (June and July), while, over the South China Sea, MTCs probability is large in early (June) and late (September) portions of the summer season. In both these regions, the preference for MTCs in June could be related to the fact that the ITCZ is closer to the equator. Thus there is a more significant influence of the cross-equatorial low-level jet on systems born in this period.

\noindent Figure~\ref{fig:FIG5}d,e, and f show the latitudinal genesis and lysis probability distribution (in \%) over the Arabian Sea, Bay of Bengal, and the South China Sea. In the latter two regions, the mean genesis locations are equatorward compared to lysis latitudes. This is consistent with the noted northwestward propagation tendency of these systems. On the other hand, the genesis and lysis distributions over the Arabian Sea largely coincide, which is in agreement a quasi-stationary character or a slow westward motion. Furthermore, there is a peak in these systems' genesis over all three basins at approximately $16^{\circ}$N. 
 
\noindent Figures~\ref{fig:FIG5}g,h show distributions of $\delta\xi_{p}$ and $\xi_m$ with latitude, respectively. As indicated earlier, the MTC-phase probability is higher towards the equator and this is also seen in Figure~\ref{fig:FIG5}g. In fact, equatorward of 15$^\circ$N, almost all the systems center show $\delta \xi_p > 0$. On the other hand, Figure \ref{fig:FIG5}h suggests that cyclonic centers tend to become stronger with latitude (up to about 21$^\circ$N).
In other words, taken together, Figures~\ref{fig:FIG5}g,h indicate that away from equator systems are more intense and less localized in the middle tropospheric (LTC-phase). Whereas, near the equator, they are less intense but more localized in the middle troposphere (MTC-phase).

%=======================================================
%                MTCS OVER THE GLOBE
%=======================================================
\subsection{MTC statistics over the Globe}
\subsubsection{Boreal Summer}
In the previous section, cyclone statistics over South Asia, including track density, genesis lysis density, the preferred direction of motion and monthly frequency, were discussed. In particular, we noted that systems classified as MTCs, i.e., they showed a middle tropospheric vorticity maximum for at least two days, can change character. Indeed, most of these type systems can be viewed as having two phases: an MTC-phase and LTC-phase. Keeping this in mind, here, we focus on the boreal summer. Moreover, we perform an analysis of over $30^{\circ}$N-$30^{\circ}$S to identify potential areas that support MTCs and LTCs. 
We note that imposing multiple criteria to filter and track synoptic phenomena can result in a fair amount of inconsistency among different reanalysis products \citep{hodges2003comparison}. Further, as we are now moving to a global perspective, manual tracking proves to be extremely time intensive. 
Since our intention here is to identify regions that support MTCs or LTCs with composite structures --- hence, the center-density suffices for our purpose. In all, for the 20 years, daily 600 hPa geopotential surfaces are examined during the boreal summer. As we do not have tracks, we cannot trace the MTC and LTC phases of a system; instead, we get a quantitative snapshot of mid and lower tropospheric cyclonic center density in the boreal summer.
Note that for systems in South Asia, almost 86\% MTCs identified were moist. Given that the precipitable water in the tropics varies with longitude \citep{Sukhatme1}, it is essential to enforce the moisture constraint when considering systems from different regions of the globe. 
 
\begin{figure*}
\centering
\includegraphics[trim=0 0 0 0, clip,height = 1.0\textwidth,width = 1.0\textwidth, angle =0, clip]{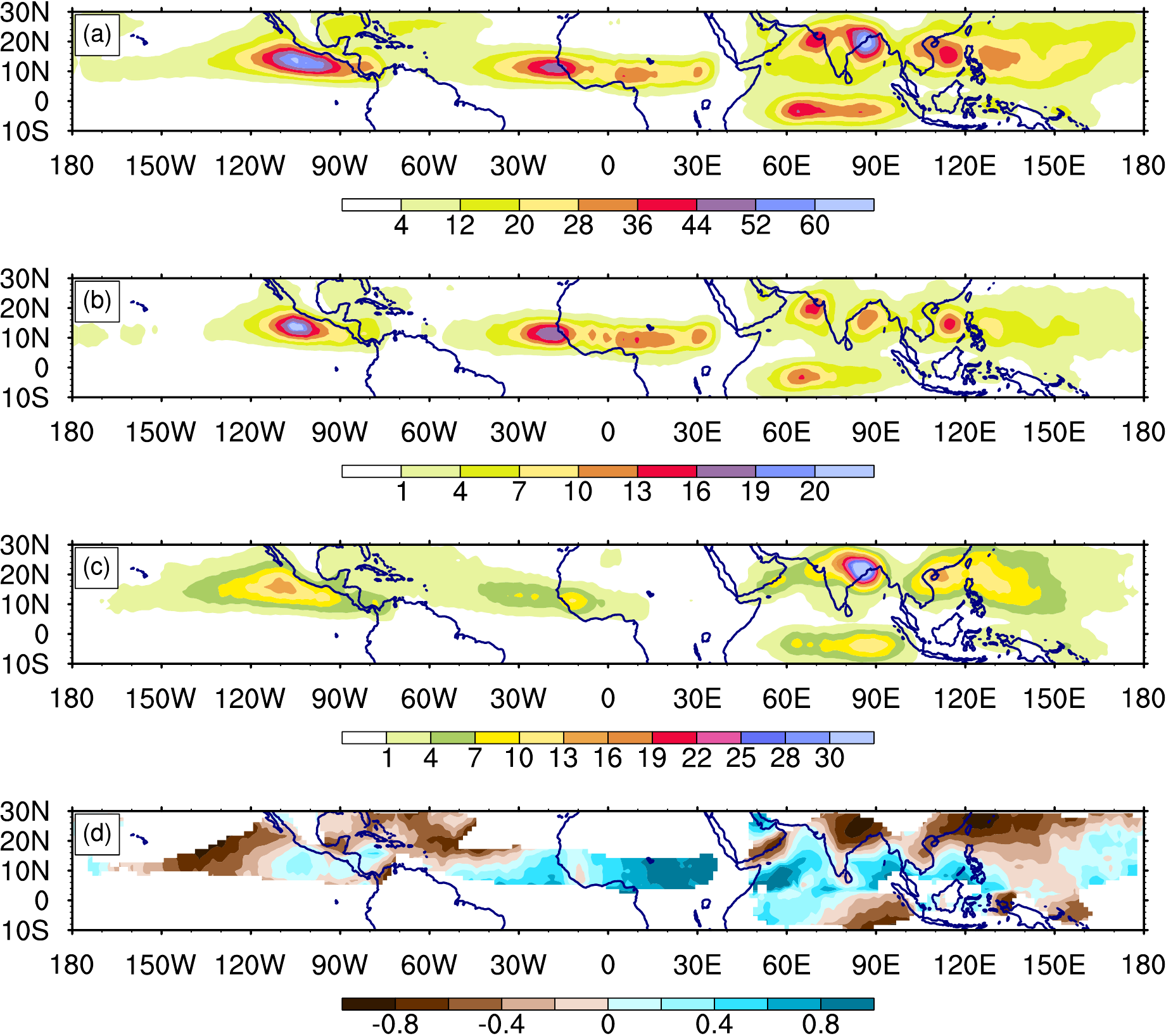}
\caption{(a) Overall cyclone center density %collected over 8-degree square at 1.5 degree grid per 
for Northern Hemisphere summer (i.e., June, July, August, September) for 20 years; (b) same as (a) except for MTC centers %($\delta \xi_{p}>1.5$, $650 \geq P_{\xi} \geq 500$ hPa, $\xi_{m}>1.5$, $\xi_{l}>0$, $Q_{m}>2$); 
(c) same as (a) except for LTC centers %($\delta \xi_{p}<0.5$, $700 \leq P_{\xi} \leq 1000$ hPa,  $\xi_{m}>1.5$, $\xi_{l}>0$,  $Q_{m}>2$), i.e., LTCs; 
(d) $\frac{b-c}{b+c}$, only shown if total density exceeds 3; Unit of vorticity is $10^{-5}s^{-1}$ and specific humidity is in $g/kg$}
\label{fig:FIG6}
\end{figure*}

\noindent In all, 49,580 non-topographic moist cyclonic systems with $\xi_m > 1.5 \times 10^{-5}$s$^{-1}$ were detected in the boreal summer over 20 years. 
The cyclone center density of all these systems is shown in Figure~\ref{fig:FIG6}a. Out of these, 10,406  
centers satisfy MTC criteria, and 11,984 are of LTC type. Again, we aim to isolate systems with middle and lower tropospheric characteristics, so about 50\% of the detected centers do not fall in either the MTC or LTC category. These filtered subsets are used to estimate the cyclone density maps of LTCs and MTCs in Figure~\ref{fig:FIG6}b and c. Taken together, Figure~\ref{fig:FIG6} suggests that
over South Asia, regions of dominant cyclonic activity include the Bay of Bengal ($16^{\circ}$N-$20^{\circ}$N, $85^{\circ}$E-$90^{\circ}$E), South China Sea ($16^{\circ}$N-$22^{\circ}$N, $110^{\circ}$E-$130^{\circ}$E), and the Arabian Sea ($18^{\circ}$N-$22^{\circ}$N, $68^{\circ}$E-$74^{\circ}$E), which are consistent with the manual tracking used in the previous section (Figure~\ref{fig:FIG4}). Further, two maxima are evident over Africa (West and East-Central Africa), one over the East Pacific ($5^{\circ}$N-$15^{\circ}$N, $120^{\circ}$W-$90^{\circ}$W) near Central America. The maximum over the South China Sea extends and connects with a second maximum over the West Pacific. Finally, another prominent region of cyclone activity lies just south of the equator over the Indian Ocean. This region matches the zone of maximum rainfall south of the equator during June-September \citep{adler2017global}. 
Figures~\ref{fig:FIG6}b,c, and d suggest that MTC fraction increases near the equator, while LTC fraction increases as we move towards the subtropics.

\noindent The total cyclone center density over South Asia is consistent with previous regional studies of cyclonic activity \citep{Raymond,HurleyBoos, Boos2015,Hunt}. As expected from the metric's nature, the cyclone density maxima here are situated in regions where cyclones show minimum speed. Figure~\ref{fig:FIG6}d suggests that most of the cyclones detected over the westy Pacific and northeast India are LTCs, which is consistent with the relatively high frequency of monsoon lows in these regions \citep{HurleyBoos}.
Moreover, over the Arabian Sea, 
consistent with the previous section, southwards of $20^{\circ}$N, there is a higher propensity of middle troposphere systems, and lower tropospheric systems are prominent as one progresses northwards. 

\noindent Cyclone activity over Africa
is high over the west coast ($5^{\circ}$N-$15^{\circ}$N $30^{\circ}$W-$5^{\circ}$W) and over the east-central continent ($5^{\circ}$N-$20^{\circ}$N, $10^{\circ}$E-$40^{\circ}$E) as seen in Fig~\ref{fig:FIG6}a, both of which are consistent with regional studies \citep{kiladis2006three}. These regions, during June-September, receive almost 95\% of their annual rainfall \citep{mohr2004interannual} and are characterized by a low-level westerly-jet, large north-south temperature gradient, middle (850 hPa-600 hPa), and upper level (300 hPa-150 hPa) easterly jet. 
Here, the movement of weather systems is primarily controlled by the African Easterly Jet (AEJ).
Usually, cyclonic disturbances are driven to the west coast of Africa and
some of them spin-up at lower levels to become warm-core tropical lows \citep{hopsch2010analysis}. 
Further, it has been suggested that African Easterly Waves (AEWs) can be induced downstream by such disturbances near the entrance of the AEJ \citep{kiladis2006three,thorncroft2008three}. Thus, the high cyclone center density observed near the AEJ entrance in Figure~\ref{fig:FIG6} over east Africa may be due to these convective systems. The second maximum, over west Africa, is likely a reflection of the downstream induced systems embedded in or are part of AEWs \citep{thorncroft2008three}.  

\noindent A comparison of MTC and LTC densities in Figure~\ref{fig:FIG6}d suggests that over the ocean, off the west coast of Africa, LTCs are more abundant as compared to East and Central Africa.
Whereas the MTC fraction becomes progressively higher eastwards over African landmass. 
In fact, recent studies suggest that troughs of AEWs over East and Central Africa are usually rich in stratiform convection, thereby enhancing the middle troposphere maximum of relative vorticity \citep{russell2020african,russell2020potential}.
Hence, the large fraction of MTCs over these regions may reflect top-heavy latent heating from stratiform clouds. More broadly, it should be noted that many regions worldwide do not show the presence of systems analogous to AEWs; however, they do support MTCs. This suggests that the MTC-phase is not an inherent property of AEWs. 
The larger number of LTCs off Africa's coast is consistent with the transformation of some AEWs into tropical lows or even cyclones over the ocean \citep{hopsch2010analysis}. Further, \cite{russell2020potential} suggests western Africa and the Atlantic Ocean are rich in deep convection, thus enhancing the low-level circulation of systems in these regions. 

\begin{figure*}
\centering
\includegraphics[trim=0 0 0 0, clip,height = 1\textwidth,width = 0.5\textwidth, angle =0, clip]{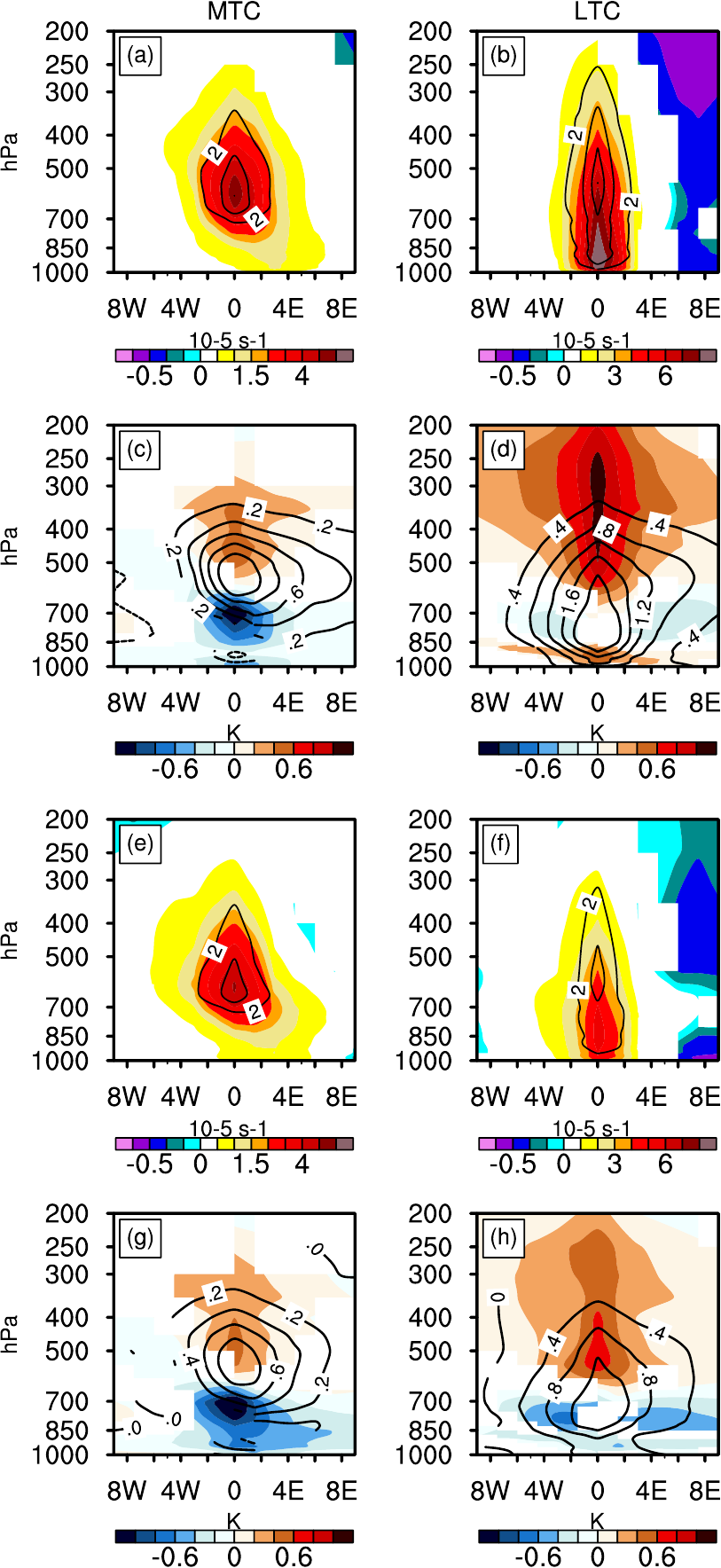}
\caption{Composite of MTCs (left column) and LTCs (right column) for the East Pacific region (a,b,c,d) and the Africa (e,f,g,h); (a,b) Relative vorticity anomaly and contours of potential vorticity anomaly; (c,d) temperature anomaly and specific humidity anomaly in contours. Results are only shown if they are significant at 99\% confidence under two tailed t-test. }
\label{fig:FIG7}
\end{figure*}

\noindent Composites of middle and lower tropospheric systems over Africa are shown in Figure \ref{fig:FIG7}e,f,g, and h. MTCs clearly have a vorticity maximum in the middle troposphere and baroclinic thermal structure that shows remarkable similarity with systems detected over South Asia (Figure \ref{fig:FIG3}). Besides, the vorticity and moisture anomalies extend downwards but relatively east of the center, reflecting an east-west tilt of the system.    
The PV anomalies of LTCs and MTCs are distinct, with the latter being confined in the middle troposphere. At the same time, the former extends down to the lower troposphere maintaining an upright structure. 
Finally, 
we obtain a low cyclone density near the Sahara desert, which contrasts with previous studies that suggested high activity over north and northwest of Africa \citep{HurleyBoos}. As discussed earlier, this difference is due to the use of 600 hPa level for identification, which does not account for shallow, near-surface lows.

\noindent Also, consistent with case studies \citep{ReedR,Shapiro}, another geographical region of significant moist mid-level cyclonic activity is in the eastern Pacific off the coast of Central America ($120^{\circ}$W-$90^{\circ}$W and $0^{\circ}$N-$15^{\circ}$N). This region was also identified in the recent survey of low-pressure systems by \cite{HurleyBoos}. Here, the MTCs (LTCs) are favored close to land (over the open ocean).  Indeed, the fractional density (Figure~\ref{fig:FIG6}d) changes from positive values between 90$^\circ$W-120W$^\circ$ (MTC) to negative values over $120^{\circ}$W-$150^{\circ}$W (LTC). A composite made from systems in this region is shown in Figure \ref{fig:FIG7}a,b,c, and d. The distinction between the two types of cyclonic centers is clear. Further, as was the case with Africa, the MTC category is structurally similar to middle tropospheric systems over South Asia (Figure \ref{fig:FIG3}a,b,c,d). A notable difference is that here the anomalies are relatively stronger. 
In fact, cyclone activity in this region shares some interesting features with the west African maximum. In particular, the closeness to the equator, the fraction of MTCs and LTCs, and the fact that both areas show higher MTC (LTC) density southeast (northwest) of the maximum in total cyclone density. Besides, both the regions are dominated by low level negative absolute vorticity advection \citep{tomas1999influence}.
Thus, in addition to mid-level cloud heating \citep{choudhury2018phenomenological,russell2020african,russell2020potential}, in these regions, the MTC profile might be aided by the lowering of low-level relative vorticity via cross hemisphere absolute vorticity advection. 

\subsubsection{Boreal Winter}
\begin{figure*}
\centering
\includegraphics[trim=0 0 0 0, clip,height = 1.\textwidth,width = 1.0\textwidth, angle =0, clip]{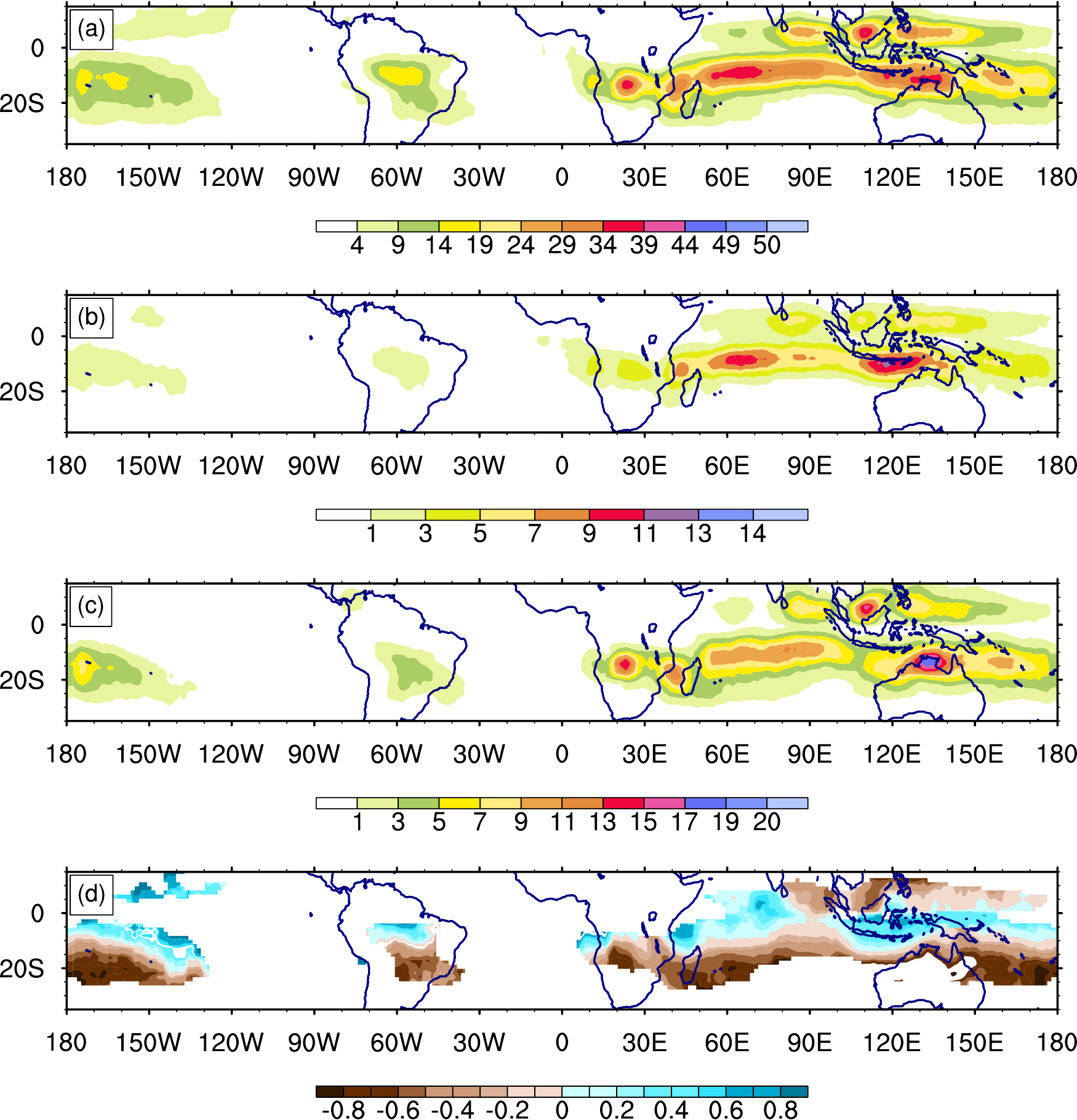}
\caption{(a) Overall cyclone center density %collected over 8-degree square at 1.5 degree grid per 
for the boreal winter (i.e., December, January, February, March) for 20 years; (b) same as (a) except for MTC centers %($\delta \xi_{p}>1.5$, $650 \geq P_{\xi} \geq 500$ hPa, $\xi_{m}>1.5$, $\xi_{l}>0$, $Q_{m}>2$); 
(c) same as (a) except for LTC centers %($\delta \xi_{p}<0.5$, $700 \leq P_{\xi} \leq 1000$ hPa,  $\xi_{m}>1.5$, $\xi_{l}>0$,  $Q_{m}>2$), i.e., LTCs; 
(d) $\frac{b-c}{b+c}$, only shown if total density exceeds 3; Unit of vorticity is $10^{-5}s^{-1}$ and specific humidity is in $g/kg$}
\label{fig:FIG8}
\end{figure*}

\begin{figure*}
\centering
\includegraphics[trim=0 0 0 0, clip,height = 1\textwidth,width = 0.5\textwidth, angle =0, clip]{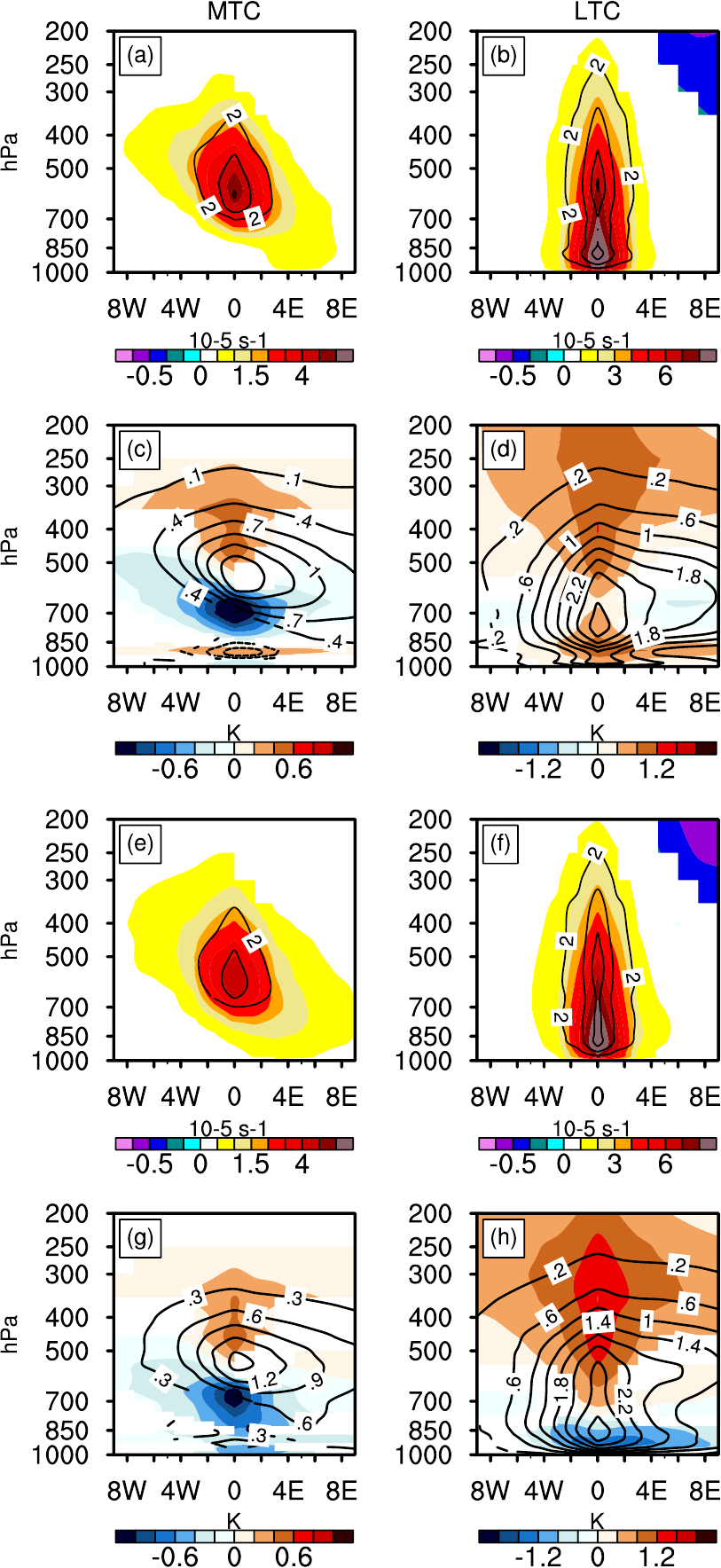}
\caption{Composite of MTCs (left column) and LTCs (right column) for the Southern Indian Ocean (a,b,c,d) and Australia (e,f,g,h); (a,b) Relative vorticity anomaly and contours of potential vorticity anomaly; (c,d) temperature anomaly and specific humidity anomaly in contours. Results are only shown if they are significant at 99\% confidence under two tailed t-test.}
\label{fig:FIG9}
\end{figure*}

The above discussion of boreal summer activity shows that MTCs are not uncommon and are observed over most monsoonal regions of the tropics  \citep{trenberth2000global}. In this section, we focus on the boreal winter (December to March; 2000-2019). Following the same methodology,
in 20 years, we find a total of 41,703 cyclonic centers during the boreal winter. 
Out of 41,703 centers, our strict isolation of middle and lower tropospheric maxima yields 6,954 systems that satisfy MTC criteria, while 10,603 belong to the LTC category. Note that we can relax our constraints to include more systems in the two categories, but to keep consistency with the analysis so far, we choose to use the same thresholds. Indeed, we still have a reasonably large sample size due to the number of years studied. These subsets are used to calculate the cyclone density shown in Figure~\ref{fig:FIG8}. 
The dominant regions of cyclonic activity turn out to be North Australia ($20^{\circ}$S-$5^{\circ}$S, $120^{\circ}$E-$145^{\circ}$E), the Southern Indian Ocean ($20^{\circ}$S-$5^{\circ}$S, $50^{\circ}$E-$90^{\circ}$E), South Africa ($20^{\circ}$S-$10^{\circ}$S, $10^{\circ}$E-$40^{\circ}$E), and subtropical South America ($25^{\circ}$S-$5^{\circ}$S, $50^{\circ}$W-$40^{\circ}$W). 
The former two show a relatively high center density as compared to the latter two regions. As with the boreal summer, here too, these regions correspond to the monsoon region of the southern hemisphere summer \citep{trenberth2000global}. Also, by and large, the above areas of cyclone activity agree with previous work on monsoon lows and moist mesoscale systems \citep{berry2016dynamics}. A notable difference is near Australia, where previous work highlighted most activity over the south-west part of the continent \citep{HurleyBoos}. This difference is again due to our use of 600 hPa field as a marker, as opposed to 850 hPa vorticity \cite[used by][]{HurleyBoos} which is strongly influenced by surface heat lows. 
As in \cite{HurleyBoos,hodges2017well}, the lowest system density is observed over South America, which indicates that most of the rainfall over there might be attributed to small scale systems that are not resolved in reanalysis or are due to open troughs, which may not be detectable on geopotential surfaces as closed contour pressure minima. Composites of middle and lower tropospheric centers from the South Indian Ocean and Australia are shown in Figure \ref{fig:FIG9}. Though the MTCs are quite similar among different regions, the LTC structure does vary. Specifically, in the South Indian Ocean, LTCs show almost no sign of a cold anomaly at low levels, and PV anomalies have a particularly clear bimodal signature. 

\noindent The highest center density seen over north Australia is consistent with the active monsoon region and consists of the westward propagating low-pressure systems that are quite similar to monsoon lows in the Indian region \citep{HurleyBoos,clark2018rainfall}. 
The region of maximum activity in North Australia extends out to the  South Pacific and merges with the South Pacific Convergence Zone \citep[SPCZ,][]{widlansky2011location}. A relatively weak signature of cyclonic activity is also found north of the equator, which corresponds to the northeast Indian monsoon \citep{rajeevan2012northeast}. Indeed, Southern India receives most of its rainfall in this season from low-pressure systems which form over the Bay of Bengal and then move westward \citep{singh2017north}. The massive rain event of 2015 that resulted in severe flooding in the state of Tamil Nadu is an example of such a weather system \citep{phadtare2018role}. Note that this region sees both lower and middle tropospheric systems (Figure~\ref{fig:FIG8}c,d). 

\noindent It is interesting to note that the LTC density is high over North Australia and is situated relatively poleward compared to the largest MTC density. Similarly, high LTC density over the Indian ocean is poleward compared to the large MTC density region.
Indeed, Figure~\ref{fig:FIG8}d is similar to its boreal summer northern hemisphere counterpart in that fractional LTC density increases away from the equator, and the MTC fractional density is relatively high towards the equator. Thus, the latitudinal preference of MTCs and LTCs is independent of the hemisphere. A common thread is that MTCs are prominent in regions where the seasonal cross-equatorial flow (advecting oppositely signed vorticity at low levels) meets with the monsoon trough LTCs are observed further northward in the vicinity of the monsoon trough itself. 

\begin{figure*}
\centering
\includegraphics[trim=0 0 0 0, clip,height = 0.8\textwidth,width = 1\textwidth, angle =0, clip]{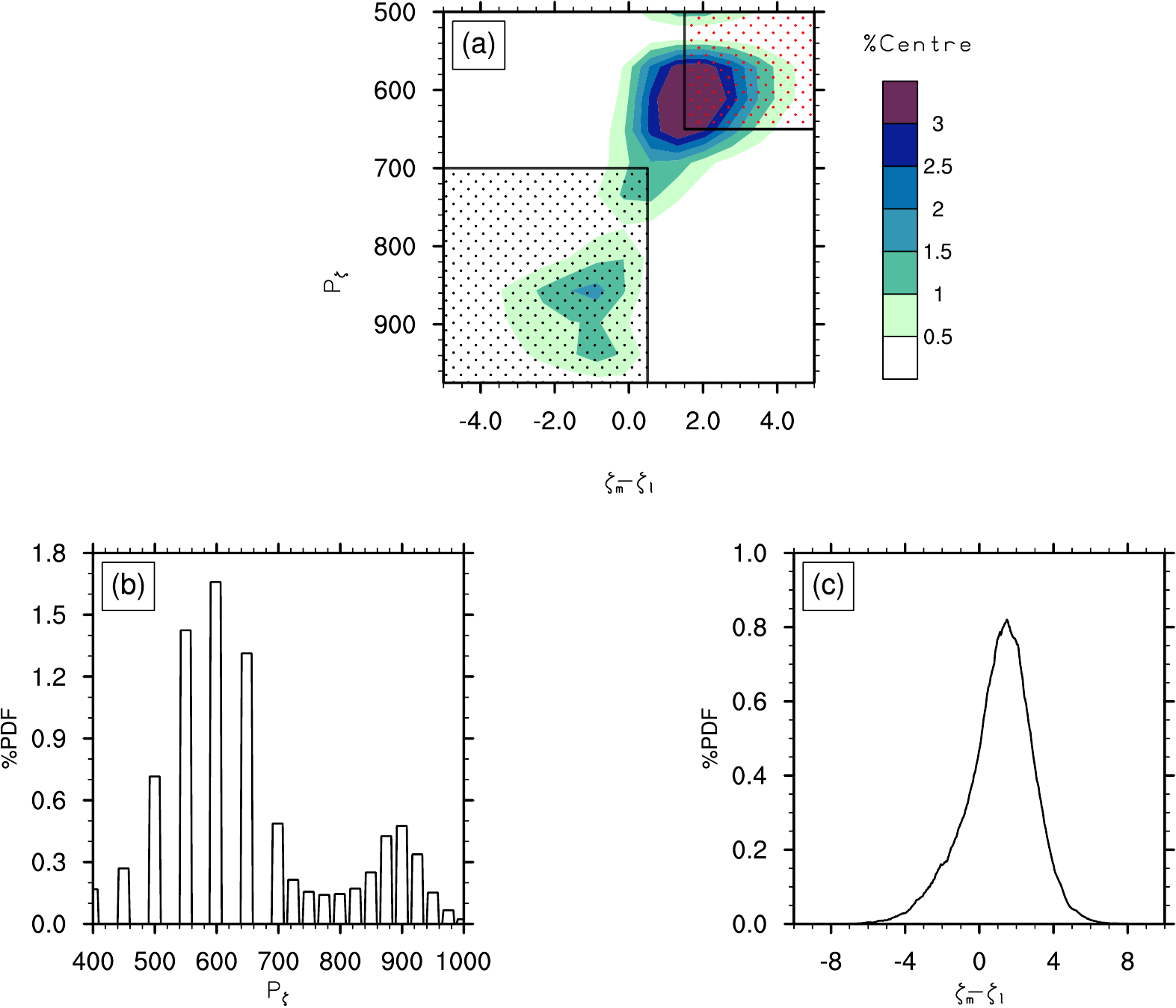}
\caption{(a) Joint PDF of $P_{\xi}$ and $\xi_{m}- \xi_{l}$ of all 91,283 cyclonic centers from $30^{\circ}$N to $30^{\circ}$S in 20 years of data. Notation follows Figure \ref{fig:FIG1}. (b) PDF of $P_{\xi}$; (c) PDF of $\xi_{m}-\xi_{l}$.}
\label{fig:FIG10}
\end{figure*}

\noindent In all, taking the boreal summer and winter together, we identified 91283 moist cyclonic centers with $\xi_m > 1.5 \times 10^{-5}$s$^{-1}$ for the entire tropics. A joint pdf of $P_\xi$ and $\xi_m -  \xi_l$ using all these systems is shown in Figure \ref{fig:FIG10}a and the distribution of $P_\xi$ itself is shown in Figure \ref{fig:FIG10}b. Both these plots highlight the bimodal nature of moist tropical systems. Indeed, cyclonic centers have clear peaks that align with the LTC or MTC categories. This also alleviates concerns about using our criteria that selected only about 50\% of the systems in the boreal winter and summer composites. It is evident from Figure \ref{fig:FIG10}b that centers are not counted either as MTCs or as LTCs mainly because $P_\xi$ varies from near surface to about 400 mbar, and we select particular layers from this distribution for each category.

\subsection{Discussion} 
At this stage, it is worthwhile to inquire into the robustness of the results and whether the identified MTCs are artifacts of the reanalysis or whether they represent the real atmosphere. Although reanalysis post-2000 performs satisfactorily in detecting cyclones \citep{hodges2017well}; arguably, the low spatial resolution of the underlying model of these products may lead to a cold-core lower troposphere structure and an under resolved lower tropospheric circulation \citep{manning2007evolution}. The first clue that these are not due to reanalysis errors is that the geographical distribution of middle tropospheric activity is in agreement with reported MTC locations of the IMD \citep{choudhury2018phenomenological} and with the Indian Ocean field experiment over the Arabian Sea \citep{miller1968iioe}. 
Even in other tropical regions, there is evidence for moist systems with mid-tropospheric vorticity maxima. For example, during the GATE-III experiment \citep{reed1977structure} and other observational case studies over Africa \citep{kiladis2006three}, near Central America \citep{simpson1967study,ReedR}, the Western Atlantic \citep{Shapiro} and the Eastern Pacific regions \citep{Raymond_ep}. 
Indeed, the fact that reanalysis shows mid-tropospheric vorticity maxima in the same regions that these field experiments noted such vorticity profiles is encouraging.

\noindent In addition to this observational support, we note that reanalysis shows systems with lower and mid-tropospheric maxima in the same regions. The MTC fraction increases towards the equator, but both systems are observed in similar locations throughout the tropics. Thus, there is no outright bias in the product, which forces systems to have a lower level profile of a certain kind in particular regions.
It is important to note that there may be some disagreement about the exact location and temporal coherence of these systems among different reanalysis data \citep{hodges2003comparison}. Also, if these systems are mainly functions of the large scale environment, then the identified areas of MTC and LTC activity are not expected to be susceptible to reanalysis data sets except for models' inherent biases in large-scale fields. Of course, the actual number of MTCs and LTCs in different reanalysis products might differ.

\section{Summary and Conclusions}
A global climatology of Middle Troposphere Cyclones (MTCs) for boreal summer and winter, based on 20 years (2000--2019) of reanalysis data, is presented. First, we analyze cyclonic systems in the Indian region during the IMD specified dates of middle troposphere circulation events over the Arabian Sea. 
The probability distribution of the differential vorticity (mid minus lower level) versus the height of vorticity maximum naturally yields two peaks (i.e., is bimodal), thus, in a sense, justifying the separation of cyclones into mid and lower-level systems. The two peaks are used to define thresholds that allow for a separation of cyclonic centers into middle and lower tropospheric systems. The utility of the selected thresholds is validated by composites of cyclone centers which fall in the two clusters.
\noindent 
Over South Asia, cyclonic systems are tracked manually for sixteen years. This yields a total of 261 MTCs, or about 3-4 systems per month in the summer season. The principal conclusions from an analysis of these systems are,

\begin{enumerate}
    \item The highest density of MTCs is over the North-East Arabian Sea, followed by the Bay of Bengal and the South China Sea. Further, MTCs are mostly found to occur during the early phase of monsoon (i.e., June, July) over the Arabian Sea and Bay of Bengal.
    
    \item MTCs are part of the life cycle of tropical cyclonic systems. In particular, tracks of systems identified as MTCs (i.e., they show a mid-tropospheric vorticity maximum for at least two days), change  character, and usually exhibit a middle troposphere vorticity maximum (MTC-phase) and a lower troposphere maximum (LTC-phase) during different periods of their life.  
    
    \item Cross basin motion suggests that MTCs observed in a given region can have a nonlocal or remote genesis. For example, early in the summer season, Arabian Sea MTCs appear to be local or related to the monsoon's onset. However, later, MTCs show signs of being born as low pressure systems to the east and then taking on a middle tropospheric character later in their life cycle as they move into the Arabian Sea region.
    
    \item Cyclone motion statistics suggest that MTCs show slow westward motion (or remain quasi-stationary) over the northeast Arabian Sea compared to the Bay of Bengal and the South China Sea. In fact, neither the MTC phase nor the LTC phase shows any preferred direction of motion over the Arabian Sea. 
    
    \item Analysis of the difference of middle and lower troposphere vorticity (i.e., $\delta\xi_{p}$) and mean middle tropospheric vorticity ($\xi_{m}$) indicates that the MTC (LTC)-phase is relatively weak (strong) and prevalent south (north) of $20^{\circ}$N.

\end{enumerate}
 
\noindent From South Asia, we move to the global tropics (30$^\circ$S--30$^\circ$N). Here, analyses of cyclone center density show that MTCs form over several monsoon regions of the world, both in the boreal summer and winter. Like the South Asian region, the probability density function of differential vorticity versus the height of peak vorticity of cyclonic centers in the boreal summer and winter for the global tropics is also bimodal. Indeed, it yields two distinct centers of action, one in the middle and another in the lower troposphere. The dominant regions of relatively high MTC activity are the Arabian Sea, East, West Africa, East Pacific, North Bay of Bengal, and the South China Sea in the boreal summer and North Australia, South Indian Ocean, South America, and South Africa during boreal winter. In both hemispheres, MTCs are more prevalent compared to LTCs as one moves towards the equator, while LTCs are more prominent compared to MTCs as one moves towards the subtropics. Indeed, MTCs are dominant equatorward of the monsoon trough and LTCs are in the majority in the vicinity of the monsoon trough itself. 

\noindent Composites from South Asia, Africa, the Pacific, Australia, and the South Indian Ocean show that, apart from their tilt, which is somewhat region-dependent, MTCs are remarkably similar throughout the tropics. In particular, they have distinct middle tropospheric vorticity, potential vorticity, and moisture anomaly maxima. Besides, MTCs show a baroclinic, warm above the deep cold-core, east-west tilted temperature anomalies. Though they have maximum intensity in the middle troposphere, in most places, MTCs also show a weak trough-like surface signature to the east of the middle-level center. LTCs, on the other hand, show a maximum vorticity and moisture anomaly in the lower troposphere with a shallow cold-core below 800 hPa and a relatively warm upright temperature structure. The PV anomalies of MTCs are unimodal, with the middle troposphere maximum.
In contrast, the PV of LTCs is usually bimodal, with one peak in the middle troposphere and another in the lower troposphere. Further, LTCs show some regional structural differences, primarily in the low-level cold anomaly, which is almost non-existent in some regions (for example, in the South Indian Ocean and East Pacific) and somewhat deeper in other parts of the tropics.

\noindent These findings pose some fundamental questions: for example, why do certain regions show a relative preference for MTCs or LTCs, and what decides the transition between LTC and MTC phases? How do the different environmental conditions in these regions affect the dynamics of lower versus mid-tropospheric systems? 
These are important issues as they may help gain insight into the development of tropical cyclones from mid-level vortices \citep{Raymond} and why tropical cyclones and monsoon depressions are more intense over some areas of the globe. Given the remarkable structural similarity of MTCs across the tropics, it is possible that they might be maintained via similar dynamical mechanisms.  
In fact, despite high SSTs, near-equatorial regions during summer months preferentially support MTC activity over LTCs. Interestingly, during monsoon seasons, near-equatorial regions show a strong cross hemisphere flow. This flow is a westerly low-level jet and results in negative vorticity advection \citep{tomas1999influence} in the northern hemisphere. In fact, in conjunction with an upper-level easterly flow, these regions are characterized by relatively large vertical shear \citep{wang1999choice,aiyyer2006climatology}. 
While top-heavy stratiform diabatic heating may enhance middle troposphere maxima \citep{choudhury2018phenomenological,russell2020african},
negative vorticity advection at low levels in conjunction with the strong vertical shear could also play a role in localizing cyclonic vorticity in the middle troposphere and supporting MTC profiles. 
Indeed, the dynamics of MTCs in favorable environments identified here are the subject of ongoing efforts.

\noindent {\it Acknowledgement:} Authors sincerely appreciate the valuable suggestions of three anonymous reviewers. PK would like to thank the Divehca Centre, IISc for financial support. JS would like to acknowledge support from University Grants Commission under project F 6-3/2018 in the Indo-Israel Joint Research Program (4$^{th}$ cycle) as well as Department of Science and Technology (DST/CCP/NCM/75/2017) under their Climate Change Programme.

\clearpage

\begin{center}
{\Large {\bf Supplementary Material}}
\end{center}

\section{Cyclone tracks with different temporal resolution 
data }
\begin{figure*}
\centering
\includegraphics[trim=0 0 0 0, clip,height = 0.9\textwidth,width = 0.9\textwidth, angle =0, clip]{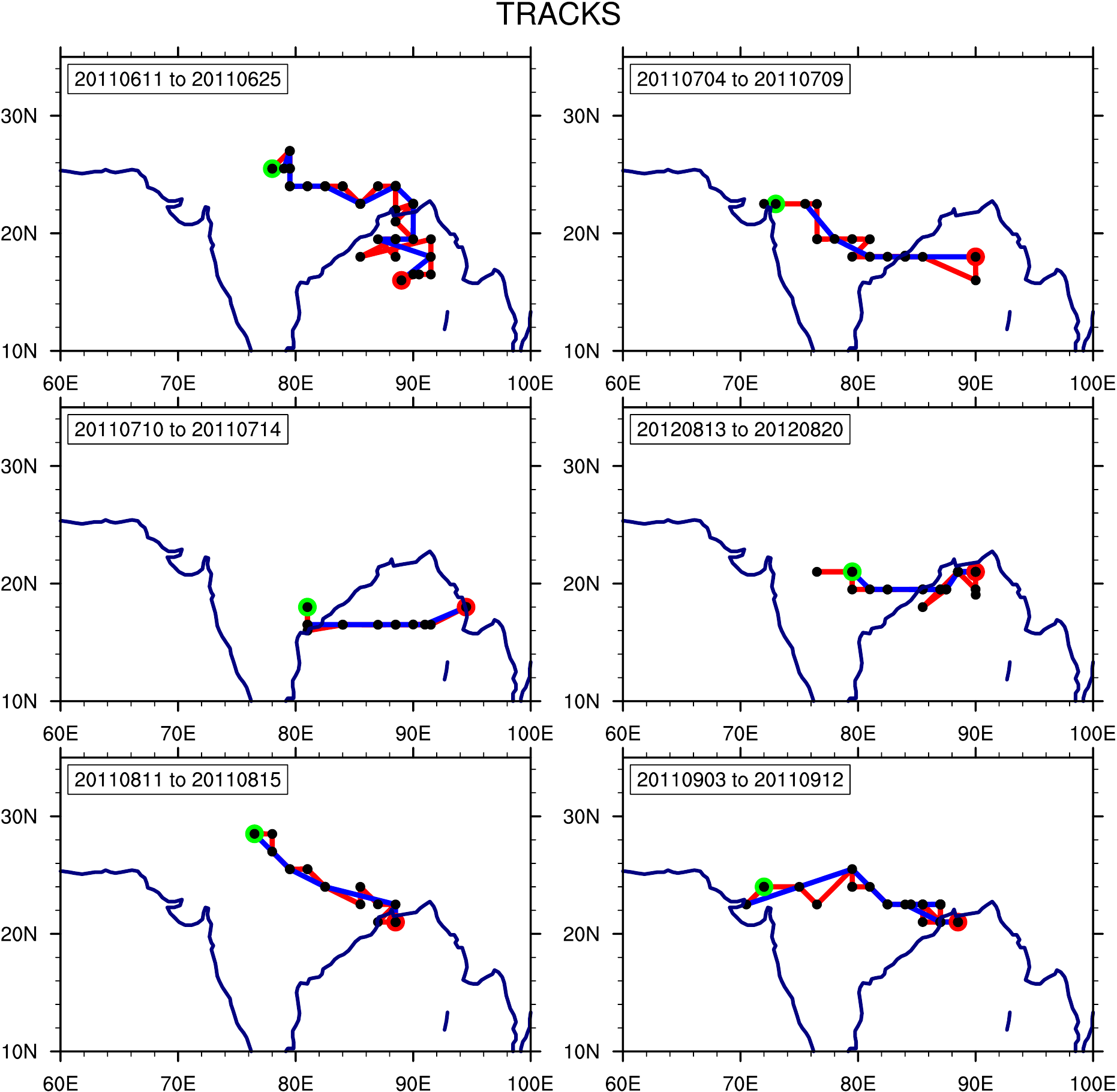}
\caption{Cyclone tracks with six hourly sampling (red) and with 24 hour sampling (blue). Red dots indicate genesis location and green dot denotes lysis; black dots represents the 6 hourly position of cyclone.}
\label{fig:FIG_S0}
%FIGURES/SUPPLYMETRY_FIGURE/SIX_PANEL_6h_24H_TRACK.pdf}
\end{figure*}

Six case studies of cyclone tracks presented are in Figure-\ref{fig:FIG_S0}. These systems are tracked manually using 6 hourly (red) and daily data (blue). As is seen, the tracks are very similar, and there is very little difference in the genesis and lysis locations of the systems. In all, apart from being smoother, some differences are apparent when cyclones abruptly change their direction of motion.

\section{Verification and sensitivity of thresholds}

We use the 725 detected non-topographic cyclonic centers from IMD MTC dates, within $5^{\circ}$N to $25^{\circ}$N and $50^{\circ}$E to $95^{\circ}$E during 1988--2008, to test the sensitivity of MTC and LTC detection as well as to verify if our thresholds can differentiate between these two types of systems. 
\subsection{Layer combinations and thickness}

\begin{figure*}
\centering
\includegraphics[trim=0 0 0 0, clip,height = 0.6\textwidth,width = 0.6\textwidth, angle =0, clip]{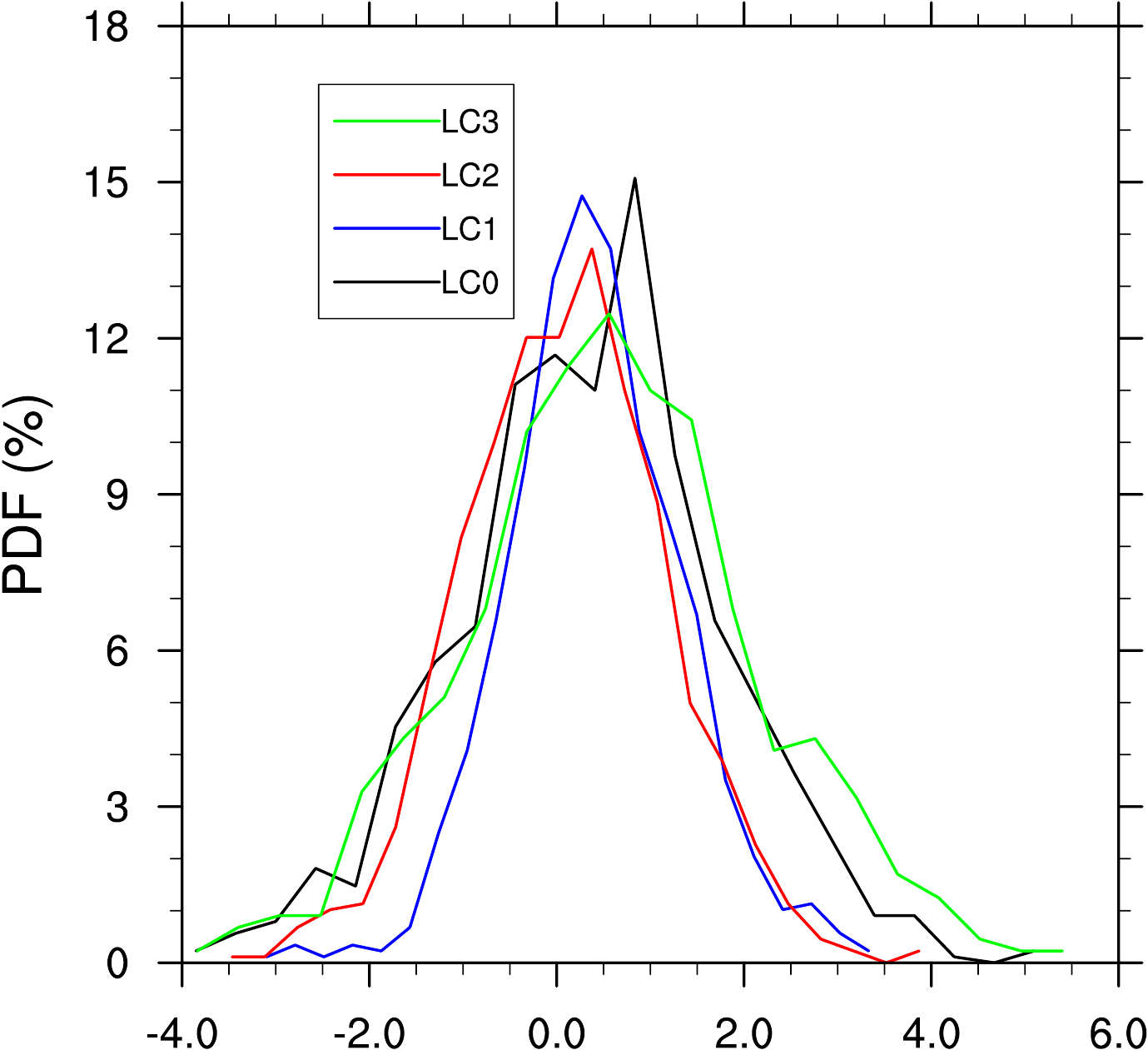}
\caption{Sensitivity of the distribution of differential vorticity ($\delta\xi_{p}$) to different choices of layers combinations. LC0: lower layer 1000-850 hPa, middle layer, 650-500 hPa; LC1: lower layer 1000-850 hPa, middle layer, 850-500 hPa; LC2: lower layer 1000-650 hPa, middle layer, 650-500 hPa and LC3: lower layer, 900 hPa, middle layer, 600 hPa.}
\label{fig:FIG_S1}
%FIGURES/SUPPLYMETRY_FIGURE/FIG_S0.pdf
\end{figure*}

To understand the effect of different choices of layers on MTC and LTC selection, 
Figure~\ref{fig:FIG_S1} shows PDFs of $\delta \xi_{p}$ of these systems for various combination of layers:
\begin{enumerate}
\item Layer combination 1: Lower layer 1000-850 hPa and middle layer 650-500 hPa (LC0: black curve): This combination is used in all results presented in the study.
    
\item Layer combination 2: Lower layer 1000-850 hPa and middle layer 850-500 hPa (LC1: blue curve); broad middle layer.
    
\item  Layer combination-3: Lower layer 1000-650 hPa and middle layer 650-500 hPa (LC2: red curve); broad lower layer. 
    
\item  Layer combination-4: Lower layer 900 hPa and middle layer 600 hPa (LC3: green curve). 
\end{enumerate}
As is seen in Figure~\ref{fig:FIG_S1}, the the choice of layers does not affect the PDF to a great extent. Examining LC3, this probability distribution is quite similar to layer combination LC0. However, this choice has the highest variance of $\delta P_{\xi}$ because no layer mean is performed; hence differences of middle and lower layer vorticity turn out to be substantial (Table~\ref{Table:TS1}). 
Further, layer combinations which are not equal in thickness, i.e., LC2 and LC1, where we have relatively thick middle or lower layers, affects the number of MTCs. This is mainly because of the shrinking of PDFs towards the origin (Figure~\ref{fig:FIG_S1}). Moreover, the reduction in number of MTCs is large compared to LTCs because the MTC threshold of $\delta \xi_{p}>1.5 \times 10^{-5}$ $s^{-1}$ is near the outer region of the tail of PDF, and LTC threshold $\delta \xi_{p}<0.5 \times 10^{-5}$ $s^{-1}$ is near the origin (for numbers, see Table~\ref{Table:TS1}). Thus, it is inappropriate to use a single level (LC3) or skewed layer thicknesses (LC2, LC1), instead the combination LC0 is most appropriate.

\subsection{Differentiation between LTCs and MTCs}

Out of the 725 centers, 121 satisfy the criteria for MTCs ($\xi_{m}-\xi_{l}>1.5 \times 10^{-5} s^{-1}$, $650 \geq P_{\xi}\geq 500$ hPa and $Q_{m}>2$, $\xi_{m}>1.5 \times 10^{-5}s^{-1}$, $\xi_{l}>0$), and 246 centers satisfy the criteria for LTCs ($\xi_{m}-\xi_{l}<0.5 \times 10^{-5} s^{-1}$, $700 \leq P_{\xi} \leq 1000$ hPa and $Q_{m}>2$, $\xi_{m}>1.5 \times 10^{-5}s^{-1}$, $\xi_{l}>0$). 

\begin{figure*}
\centering
\includegraphics[trim=0 0 0 0, clip,height = 0.9\textwidth,width = 0.6\textwidth, angle =0, clip]{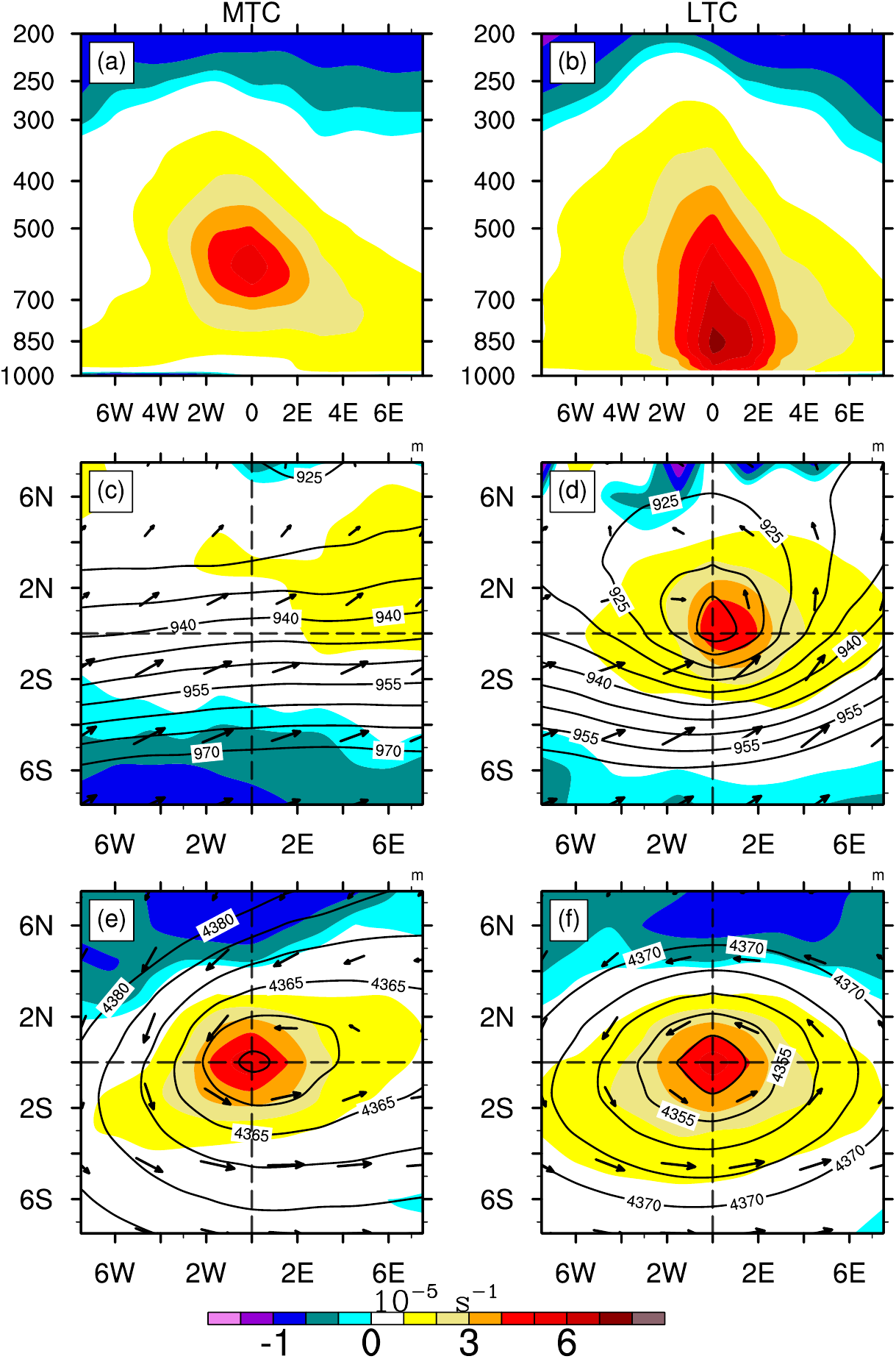}
\caption{Relative vorticity composite from 121 middle troposphere cyclonic centers (Left Panels: a, c, e) and 246 lower tropospheric cyclone centers (Right Panels: b, d, f) from the Indian region. (a), (b) East-West-vertical cross-section through composite center y-axis is pressure in hPa. (c), (d) Horizontal cross-sections at 975 hPa and (e), (f) at 600 hPa. Dashed lines indicate axes through the composite center. Vectors indicate the composite wind of MTC and LTCs and contours represent the composite geopotential height at respective levels in m. Units of the x-y axes of sub-figures c-f are degrees east and degrees north with respect to composite cyclone center, respectively.}
\label{fig:FIG_S2}
%FIGURES/FIG_M4.pdf
\end{figure*}

\noindent A composite of these LTC and MTC systems in the longitude-height plane is shown in  Figures \ref{fig:FIG_S2}a,b. For MTCs, i.e., Figure \ref{fig:FIG_S2}a, we observe a distinct mid-tropospheric vorticity maximum around 600 hPa along with weak or no signature in the lower troposphere.
On the other hand, for LTCs,  Figure \ref{fig:FIG_S2}b shows a vorticity maximum in the lower troposphere (1000-850 hPa). These features are more clearly seen in horizontal cross-sections at the 975 hPa and 600 hPa levels shown in  Figure \ref{fig:FIG_S2}c,d, and panels e,f, respectively. Both LTCs and MTCs show comparable vorticity magnitudes and closed circulation at 600 hPa. However, at 975 hPa, the MTC relative vorticity is small and slightly positive in the northeast sector with no closed circulation while the LTC composite shows comparable magnitude and closed circulations both in middle and lower troposphere. 
Thus, an overall comparison of the two categories suggests that the filtering method and the choice of threshold parameters can distinguish between middle tropospheric and lower tropospheric cyclones. 

\subsection{Moisture and relative vorticity thresholds}
Table~\ref{Table:TS1} also shows the number of MTCs and LTCs corresponding to various values of moisture ($Q_{m}$) and vorticity ($\xi_{m}$) thresholds. As expected, the number of MTCs and LTCs reduces with higher thresholds. Also, the number of LTCs is relatively less sensitive to moisture thresholds than MTCs at lower moisture thresholds. This indicates that, in general, LTCs are more moist in nature as compared to MTCs. 

\subsection{Relaxation of $P_{\xi}$ threshold}
\begin{figure*}
\centering
\includegraphics[trim=0 0 0 0, clip,height = 0.7\textwidth,width = 1\textwidth, angle =0, clip]{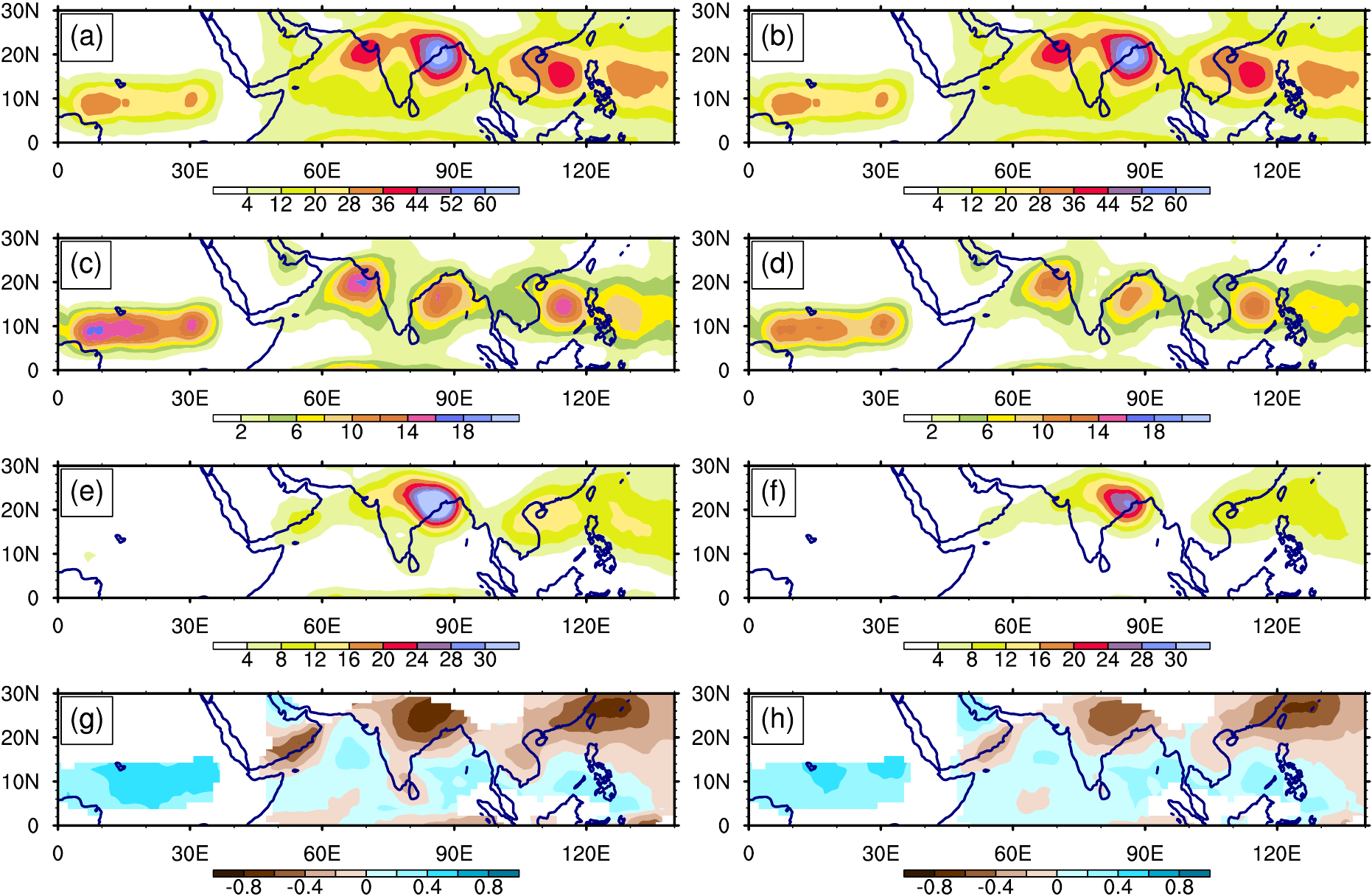}
\caption{Overall-center, LTC, MTC center density and MTC to LTC fraction from June to September 2000-2019; left panel (a, c, e, g) without $P_{\xi}$ constraints; Right panel (b, d, f, h) with all constraints, respectively.}
\label{fig:FIG_S3}
%FIGURES/SUPPLYMETRY_FIGURE/FIG_S1.pdf}
\end{figure*}

We have relaxed the constraint on the level of vorticity maxima ($P_{\xi}$) to delineate MTC and LTC phases during tracking. 
Figure~\ref{fig:FIG_S3} shows overall, MTC and LTC center density and their fraction in the presence of $P_{\xi}$ threshold and without it, respectively. As expected, the qualitative nature of results remains the same with a slight difference in MTC and LTC  activity regions.

%========================================
%========================================

\begin{table}[]
\centering
\begin{tabular}{|l|l|l|l|l|l|l|l|l|l|l|l|l|}
\cline{1-5} \cline{7-13}
\textbf{}         & \textbf{}    & \textbf{}    & \textbf{}    & \textbf{}    &  & \textbf{$Q_{m}$} & \textbf{MTC} & \textbf{LTC} & \textbf{} & \textbf{$\xi_{m}$} & \textbf{MTC} & \textbf{LTC} \\ \cline{1-5} \cline{7-13} 
\textbf{}         & \textbf{LO0} & \textbf{LO1} & \textbf{LO2} & \textbf{LO3} &  & 1.0        & 145          & 251          & \textbf{} & 1.0           & 121          & 290          \\ \cline{1-5} \cline{7-13} 
\textbf{MEAN}     & 0.36         & 0.41         & 0.09         & 0.57         &  & 2.0        & 121          & 246          &           & 2.0           & 120          & 203          \\ \cline{1-5} \cline{7-13} 
\textbf{VARIANCE} & 1.95         & 0.85         & 1.17         & 2.41         &  & 3.0        & 106          & 233          &           & 3.0           & 77           & 126          \\ \cline{1-5} \cline{7-13} 
\textbf{MEDIAN}   & 0.38         & 0.36         & 0.093        & 0.55         &  & 4.0        & 98           & 226          &           & 4.0           & 29           & 62           \\ \cline{1-5} \cline{7-13} 
\textbf{MTC}      & \textbf{121} & \textbf{52}  & \textbf{66}  & \textbf{157} &  & 5.0        & 96           & 223          &           & 5.0           & 6            & 35           \\ \cline{1-5} \cline{7-13} 
\textbf{LTC}      & \textbf{246} & \textbf{220} & \textbf{282} & \textbf{238} &  & 6.0        & 84           & 218          &           & 6.0           & 2            & 21           \\ \cline{1-5} \cline{7-13} 
\end{tabular}

\caption{Mean, variance, median and probability distributions of $\delta P_{\xi}$ of 882 detected systems over  $5^{\circ}$N-$25^{\circ}$N and $50^{\circ}$E-$95^{\circ}E$ (as shown in Figure~\ref{fig:FIG_S1}) and corresponding number of MTCs and LTCs for various combinations of layer thickness (left portion of the table); the number of MTCs and LTCs are highlighted in bold. The sensitivity of number of MTCs and LTCs for various moisture ($Q_{m}$) and relative vorticity ($\xi_{m}$) thresholds in in the right portion of the Table. }
\label{Table:TS1}

\end{table}
%======================================================
%======================================================  

%============================================================
%==================================================================  

%=====================================================

\newpage
\bibliographystyle{apalike}
\bibliography{ref.bib}
\end{document}